\documentclass{llncs}
\usepackage[pdftex,colorlinks=true,bookmarks=false,linkcolor=blue,citecolor=blue,urlcolor=blue] {hyperref}
\usepackage[all]{xy}
\usepackage{hhline}
\usepackage{subcaption}
\usepackage{rotating}
\usepackage{multirow}
\usepackage{amsmath,amssymb,bussproofs,epsfig,graphicx,verbatim,bm,galois,xcolor,xspace,todonotes}
\usepackage{microtype}
\usepackage{listings}
\usepackage{tikz}
\usetikzlibrary{arrows,shapes,positioning}
\usepackage[ruled,vlined,linesnumbered]{algorithm2e}
\usepackage{booktabs}

\newcommand{\circled}[1]{\raisebox{.5pt}{\textcircled{\raisebox{-.9pt} {#1}}}}
\newcommand{\ST}[1]{%
  \ensuremath{\circled{\tiny\strut \raisebox{.6ex}{#1}}}}
\newcommand*{\Scale}[2][4]{\scalebox{#1}{$#2$}}

\newcommand{\sbr}[1]{\lbrack \! \lbrack #1 \rbrack \! \rbrack}
\newcommand{\san}[1]{\ll \!\! #1 \!\! \gg\!}

\newcommand{\FeatExp}{\textit{FeatExp}}
\newcommand{\Exp}{\textit{Exp}}
\newcommand{\Stm}{\textit{Stm}}

\newcommand{\Var}{\textit{Var}}

\newcommand{\confprojname}[1]{\ensuremath{\pi_{#1}}}
\newcommand{\confproj}[2]{\ensuremath{\confprojname{#1}(#2)}}
\newcommand{\poset}[2]{\ensuremath{\langle{#1},{#2}\rangle}}

\newcommand{\impskip}[1][]{\ensuremath{\mbox{\texttt{skip}}^{#1}}}
\newcommand{\impassign}[3][]{\ensuremath{{#2} ~\mbox{\texttt{:=}}^{#1}~ {#3}}}

\newcommand{\impifdef}[3][]{\mbox{\texttt{\#if}}^{#1} ~\ensuremath{{#2}} ~{#3} ~\texttt{\#endif}}

\newcommand{\true}{\textrm{true}}
\newcommand{\false}{\textrm{false}}

\newcommand{\Kk}{\ensuremath{\mathbb{K}}}
\newcommand{\Ff}{\ensuremath{\mathbb{F}}}
\newcommand{\Aa}{\ensuremath{\mathbb{A}}}

\newcommand{\Zz}{\ensuremath{\mathbb{Z}}}
\newcommand{\Tt}{\ensuremath{\mathbb{T}}}
\newcommand{\Bb}{\ensuremath{\mathbb{B}}}
\newcommand{\Bt}{\ensuremath{\mathbb{BT}}}
\newcommand{\Bd}{\ensuremath{\mathbb{BD}}}
\newcommand{\Nn}{\ensuremath{\mathbb{N}}}

\newcommand{\Cc}{\ensuremath{\mathbb{C}}}
\newcommand{\Dd}{\ensuremath{\mathbb{D}}}

\newcommand{\SIMPLE}{\textsc{SIMPLE}}
\newcommand{\sIMPLE}{\textsc{SIMPLE}}
\newcommand{\APRON}{\textsc{APRON}}

\newcommand{\SPL}{\textsc{SPLNum$^2$Analyzer}}

\makeatletter
\newcommand{\mypm}{\mathbin{\mathpalette\@mypm\relax}}
\newcommand{\@mypm}[2]{\ooalign{%
  \raisebox{.1\height}{$#1+$}\cr
  \smash{\raisebox{-.6\height}{$#1-$}}\cr}}
\makeatother

\definecolor{darkblue}{rgb}{0,0,0.5}
\definecolor{lightgray}{rgb}{0.8,0.8,0.8}

\tikzset{
  treenode/.style = {align=center, inner sep=0pt, text centered,
    font=\sffamily},
  arn_n/.style = {treenode, circle, 
    draw=black, text width=2em},
  arn_r/.style = {treenode, circle, red, draw=red,
    text width=2em, very thick},
  arn_x/.style = {treenode, rectangle, draw=black,
    minimum width=1.25em, minimum height=1.25em}
}

\title{A Decision Tree Lifted Domain for Analyzing Program Families with Numerical Features (Extended Version)%
}

\author{ Aleksandar S. Dimovski\inst{1} \and Sven Apel\inst{2} \and Axel Legay\inst{3}
}
\institute{  Mother Teresa University, 12 Udarna Brigada 2a, 1000 Skopje, MKD 
  \\ \and
            Saarland University, Campus E1.1, 66123 Saarbrücken, Germany 
            \\ \and
            Universit\'e catholique de Louvain, 1348 Ottignies-Louvain-la-Neuve, Belgium 
            }

\begin{document}
\maketitle

\begin{abstract}
\emph{Lifted}  (\emph{family-based}) \emph{static
analysis} by abstract interpretation is capable of analyzing all variants of a program family
simultaneously, in a single run without generating any of the variants explicitly.
The elements of the underlying lifted analysis domain are tuples, which maintain one property per variant.
Still, explicit property enumeration in tuples, one by one for all variants, immediately yields combinatorial explosion.
This is particularly apparent in the case of program families that, apart from Boolean features,
contain also numerical features with big domains, thus admitting astronomic configuration spaces.

The key for an efficient lifted analysis is proper handling of
variability-specific constructs of the language (e.g.,\ feature-based runtime tests and $\texttt{\#if}$ directives).
In this work,
we introduce a new symbolic representation of the lifted abstract domain that can efficiently analyze program families
with numerical features.
This makes sharing between property elements corresponding to different variants explicitly possible.
The elements of the new lifted domain are constraint-based \emph{decision trees}, where decision nodes are labeled with
linear constraints defined over numerical features and
the leaf nodes belong to an existing single-program analysis domain.
To illustrate the potential of this representation,
we have implemented an experimental lifted static analyzer, called \SPL, for inferring invariants of C programs.
It uses existing numerical domains (e.g., intervals, octagons, polyhedra) from the \APRON\, library
as parameters.
An empirical evaluation on benchmarks from SV-COMP and BusyBox yields promising preliminary results indicating that our decision trees-based approach is effective and
outperforms the tuple-based approach, which is used as a baseline 
analysis based on abstract interpretation.
\end{abstract}

\section{Introduction}\label{sec:introduction}

Many software systems today are configurable \cite{pl-patterns-book}: they use \emph{features} (or configurable options) to control the presence and absence of software functionality. Different family members, called variants, are derived by switching features on and off, while the reuse of common code is maximized, leading to productivity gains, shorter time to market, greater market coverage, etc. Program families (e.g.,\ Software Product Lines) are commonly seen in the development of commercial embedded software, such as cars, phones, avionics, medicine, robotics, etc.
Configurable options (features) are used to either support different application scenarios for embedded components, to provide portability across different hardware platforms and configurations, or to produce variations of products for different market segments or different customers.
We consider here program families implemented using $\texttt{\#if}$ directives from the C preprocessor \texttt{CPP} \cite{KastnerPhD}.
They use $\texttt{\#if}$-s to specify under which conditions parts of code should be included or excluded from a variant.
Classical program families use only Boolean features that have two values: on and off.
However, Boolean features are insufficient for real-world program families, as there exist
features that have a range of numbers as possible values. These features are called \emph{numerical features} \cite{DBLP:conf/icse/HenardPHT15,DBLP:conf/splc/MunozOPFB19}.
For instance, Linux kernel, BusyBox, Apache web server, Java Garbage Collector represent
some real-world program families with numerical features.
Analyzing such program families 
is very challenging, due to the fact that
from only a few features, a huge number of variants can be derived.

This paper concerns the verification of program families with Boolean and numerical features
using abstract interpretation-based static analysis.
\emph{Abstract interpretation} \cite{DBLP:conf/popl/CousotC77,DBLP:journals/ftpl/Mine17} is a general theory
for approximating the semantics of programs.
It provides sound (all confirmative answers are correct) and efficient (with a good
trade-off between precision and cost) static analyses of run-time properties of real programs.
It has been used as the foundation for various successful industrial-scale static
analyzers, such as \textsc{ASTREE} \cite{DBLP:journals/fmsd/CousotCFMMR09}.
Still, static analysis of program families is harder than static analysis of single programs,
because the number of possible variants can be very large (often huge) in practice.
The simplest brute-force approach that uses a preprocessor to generate all variants of
a family, and then applies an existing off-the-shelf single-program analyzer to each individual variant, one-by-one,
is very inefficient \cite{taosd2013,DBLP:journals/tosem/RheinLJKA18}.
Therefore, we use so-called
\emph{lifted} (family-based) \emph{static analyses} \cite{taosd2013,DBLP:conf/aosd/MidtgaardBW14,DBLP:journals/tosem/RheinLJKA18}, which
analyze all variants of the family simultaneously without generating any of the variants explicitly.
They take as input the common code base, which encodes all variants of a program family, and
produce precise analysis results corresponding to all variants. They use a lifted
analysis domain, which represents an $n$-fold 
product of an existing single-program analysis domain used
for expressing program properties (where $n$ is the number of valid configurations).
That is, the lifted analysis domain
 maintains one property element per valid variant in tuples.
The problem is that this explicit property enumeration in tuples becomes computationally intractable with larger program
families because 
the number of variants (i.e.\ configurations)
grows exponentially with the number
of features. This problem has been successfully addressed for program families that contain
only Boolean features \cite{feature-interaction-apel-ase2011,DBLP:conf/icse/ApelRWGB13,spllift2013,DBLP:conf/gpce/Dimovski19}, by
using sharing through binary decision diagrams (BDDs). However, the fundamental limitation of existing lifted analysis techniques is
that they do not deal with numerical features.

To overcome this limitation, in this work we present a new, refined lifted abstract domain for effectively analyzing program
families with numerical features by means of abstract interpretation.
The elements of the lifted abstract domain are constraint-based \emph{decision trees}, where the decision nodes are
labelled with linear constraints over numerical features, whereas the leaf nodes
belong to a single-program analysis domain.
The decision trees recursively partition the space of configurations (i.e., the space of possible combinations of feature values), whereas
the program properties at the leaves provide analysis information corresponding to each partition, i.e.\  to the variants (configurations) that satisfy the constraints along the path to the given leaf node.
The partitioning is dynamic, which means that partitions are split by feature-based  tests (at $\texttt{\#if}$ directives),
and joined when merging the corresponding control flows again. In terms of decision trees, this means that
new decision nodes are added by feature-based tests
and removed when merging control flows. In fact, the partitioning of the set of configurations is semantics-based,
which means that linear constraints over numerical features that occur in decision nodes are automatically
inferred by the analysis and do not necessarily occur syntactically in the code base.

The lifted abstract domain is parametric in the choice of numerical property domain which underlies the linear constraints over numerical features labelling decision nodes,
and the choice of the single-program analysis domain for leaf nodes.
In fact, in our implementation, we also use numerical property domains for leaf nodes, which encode linear
constraints over program variables.
We use here the well-known numerical domains, such as intervals \cite{DBLP:conf/popl/CousotC77}, octagons \cite{DBLP:journals/lisp/Mine06}, polyhedra \cite{DBLP:conf/popl/CousotH78}, from the \APRON\ library \cite{DBLP:conf/cav/JeannetM09} to obtain a concrete decision tree-based implementation of the lifted abstract domain.
This way, we have implemented a \emph{forward reachability analysis} of C program families
with numerical (and Boolean) features for the automatic inference of invariants.
Our tool, called \SPL \footnote{\textsc{Num$^2$} in the name of the tool refers to its ability to both handle \textsc{Num}erical features and to perform \textsc{Num}erical client analysis of SPLs (program families).}, computes a set of possible invariants, which represent linear constraints over program variables.
We can use the implemented lifted static analyzer to check invariance properties of C program families,
such as assertions, buffer overflows, null pointer references, division by zero,
etc \cite{DBLP:conf/esop/CousotCFMMMR05}.

In summary, we make several contributions in this work:
\begin{itemize}
\item First, we propose a new, parameterized lifted analysis domain based on decision trees for analyzing program families with numerical features.
\item Then, we implement a prototype lifted static analyzer, \SPL, that performs
a forward analysis of $\texttt{\#if}$-enriched C programs, where numerical property domains from the \APRON\ library
are used as parameters in the lifted analysis domain.
\item Finally, we evaluate our approach for
automatic inference of invariants by
comparing performances of lifted analyzers based on tuples and decision trees.
\end{itemize}

\section{Motivating Example}\label{sec:motivate}

To illustrate the potential of a decision tree-based lifted domain, we
consider a motivating example using the code base of the following program family \SIMPLE:
\begin{center}
	$\begin{array}{ll}
\textcolor{gray}{{\ST 1}} \qquad & \texttt{int} \ \impassign{\texttt{x}}{10}, \, \impassign{\texttt{y}}{0}; \\
\textcolor{gray}{{\ST 2}} \qquad & \texttt{while}\, (\texttt{x !=}\, 0)\, \, \{ \\
\textcolor{gray}{{\ST 3}} \qquad & \qquad     \impassign{\texttt{x}}{\texttt{x-}1};  \\
\textcolor{gray}{{\ST 4}} \qquad & \qquad \texttt{\#if} \, (\texttt{SIZE} \leq 3) \ \impassign{\texttt{y}}{\texttt{y+}1}; \ \texttt{\#else} \ \impassign{\texttt{y}}{\texttt{y-}1}; \ \texttt{\#endif} \\
\textcolor{gray}{{\ST 5}} \qquad & \qquad \texttt{\#if} \, (!\texttt{B}) \ \impassign{\texttt{y}}{0}; \ \ \texttt{\#else} \ \texttt{skip}; \ \texttt{\#endif} \ \textcolor{gray}{{\ST 6}} \} \\
\textcolor{gray}{{\ST 7}} \qquad & \texttt{assert} \, (\texttt{y} > 1);
	\end{array}$
\end{center}
The set \Ff\ of features is $\{\texttt{B},\texttt{SIZE}\}$, where \texttt{B} is a Boolean feature and \texttt{SIZE} is
a numerical feature whose domain is $[1,4] = \{1, 2, 3, 4\}$. Thus, the set of valid configurations
is $\Kk=\{\texttt{B} \land (\texttt{SIZE}\!=\!1), \texttt{B} \land (\texttt{SIZE}\!=\!2), \texttt{B} \land (\texttt{SIZE}\!=\!3), \texttt{B} \land (\texttt{SIZE}\!=\!4),
 \neg \texttt{B} \land (\texttt{SIZE}\!=\!1), \neg \texttt{B} \land (\texttt{SIZE}\!=\!2), \neg \texttt{B} \land (\texttt{SIZE}\!=\!3), \neg \texttt{B} \land (\texttt{SIZE}\!=\!4) \}$.
The code of \SIMPLE\ contains two \texttt{\#if} directives, which change the value assigned to \texttt{y},
depending on how features from $\Ff$ are set at compile-time.
For each configuration from $\Kk$, a different variant (single program) can be generated by appropriately
resolving \texttt{\#if}-s.
For example, the variant corresponding to configuration
$\texttt{B} \land (\texttt{SIZE}\!=\!1)$ will have \texttt{B} and \texttt{SIZE} set to \true\ and 1, so that the
assignment $\impassign{\texttt{y}}{\texttt{y+}1}$ and \texttt{skip} in program locations \ST 4 and \ST 5, respectively, will be included in this variant.
The variant for configuration
$\neg \texttt{B} \land (\texttt{SIZE}\!=\!4)$ will have features \texttt{B} and \texttt{SIZE} set to \false\ and 4, so the
assignments $\impassign{\texttt{y}}{\texttt{y-}1}$ and \impassign{\texttt{y}}{0} in program locations \ST 4 and \ST 5, respectively, will be included in this variant.
There are $|\Kk|=8$ variants that can be derived from the family \SIMPLE.

Assume that we want to perform  \emph{lifted polyhedra analysis} of \SIMPLE\ using the \emph{Polyhedra} numerical
domain \cite{DBLP:conf/popl/CousotH78}.
The standard lifted analysis domain used in the literature \cite{taosd2013,DBLP:conf/aosd/MidtgaardBW14} is
defined as cartesian product of $|\Kk|$ copies of the basic analysis domain (e.g. polyhedra).
Hence, elements of the lifted domain are tuples
containing one component for each valid configuration from $\Kk$, where each component represents a polyhedra linear constraint
over program variables (\texttt{x} and \texttt{y} in this case).
The lifted analysis result in location \ST 7 of \SIMPLE\ 
is an 8-sized tuple shown in Fig.~\ref{fig:tuple}.
Note that the first component of the tuple in Fig.~\ref{fig:tuple} corresponds to configuration $\texttt{B} \land (\texttt{SIZE}\!=\!1)$, the second to
$\texttt{B} \land (\texttt{SIZE}\!=\!2)$, the third to
$\texttt{B} \land (\texttt{SIZE}\!=\!3)$, and so on.
We can see in Fig.~\ref{fig:tuple} that the polyhedra analysis discovers
very precise results for the variable \texttt{y}: ($\texttt{y} \!=\! 10$) for configurations $\texttt{B} \land (\texttt{SIZE}\!=\!1)$,
$\texttt{B} \land (\texttt{SIZE}\!=\!2)$, and $\texttt{B} \land (\texttt{SIZE}\!=\!3)$; ($\texttt{y} \!=\! -10$) for configuration $\texttt{B} \land (\texttt{SIZE}\!=\!4)$; and ($\texttt{y} \!=\! 0$) for all other configurations.
This is due to the fact that the polyhedra domain is fully relational and is able to track all relations between program variables \texttt{x} and \texttt{y}.
Using this result in location \ST 7, we can successfully conclude that the assertion is valid for configurations
$\texttt{B} \land (\texttt{SIZE}\!=\!1)$, $\texttt{B} \land (\texttt{SIZE}\!=\!2)$, and $\texttt{B} \land (\texttt{SIZE}\!=\!3)$,
whereas the assertion fails for all other configurations.

\begin{figure*}[t]
\centering
\begin{minipage}[b]{0.47\linewidth}
\centering
$\Scale[0.81]{ \begin{array}{@{} l @{}}
\!\!\Big(\,\overbrace{[\texttt{y} \!=\! 10, \texttt{x} \!=\! 0]}^{\texttt{B} \land (\texttt{SIZE}\!=\!1)}, \overbrace{[\texttt{y} \!=\! 10, \texttt{x} \!=\! 0]}^{\texttt{B} \land (\texttt{SIZE}\!=\!2)},  \overbrace{[\texttt{y} \!=\! 10, \texttt{x} \!=\! 0]}^{\texttt{B} \land (\texttt{SIZE}\!=\!3)}, \\ \overbrace{[\texttt{y} \!=\! -10, \texttt{x} \!=\! 0]}^{\texttt{B} \land (\texttt{SIZE}\!=\!4)},
\overbrace{[\texttt{y} \!=\! 0, \texttt{x} \!=\! 0]}^{\neg \texttt{B} \land (\texttt{SIZE}\!=\!1)}, \overbrace{[\texttt{y} \!=\! 0, \texttt{x} \!=\! 0]}^{\neg \texttt{B} \land (\texttt{SIZE}\!=\!2)},  \\
\overbrace{[\texttt{y} \!=\! 0, \texttt{x} \!=\! 0]}^{\neg \texttt{B} \land (\texttt{SIZE}\!=\!3)}, \overbrace{[\texttt{y} \!=\! 0, \texttt{x} \!=\! 0]}^{\neg \texttt{B} \land (\texttt{SIZE}\!=\!4)} \, \Big)
	\end{array}}$
\vspace{-1mm}
\caption{Tuple-based analysis result at location \ST 7 of \SIMPLE.}
\label{fig:tuple}
\end{minipage}
\begin{minipage}[b]{0.47\linewidth}
\centering
\begin{tikzpicture}[-,>=stealth',level/.style={sibling distance = 1.1cm/#1,
  level distance = 0.3cm}]
     \node[arn_n](A){\scriptsize{\texttt{B}}};
     \node[arn_n,below left=of A](B){\tiny{$\texttt{SIZE}\!\!\leq\!\!\!3$}};
     \node[arn_x,below right=of A](C){${\scriptstyle [\texttt{y} = 0 \land \texttt{x} = 0]}$};
     \node[arn_x,below left=1.08cm and 0.05cm of B](F){${\scriptstyle [\texttt{y} = 10 \land \texttt{x} = 0]}$};
     \node[arn_x,below right=1.08cm and 0.05cm of B](D){${\scriptstyle [\texttt{y} = -10 \land \texttt{x} = 0]}$};
     \draw[dashed](A) to node[above right]{} (C);
     \draw[-](A) to node[above left]{} (B);
     \draw[-](B) to node[above left]{} (F);
     \draw[dashed](B) to node[above left]{} (D);
  \end{tikzpicture}
\vspace{-1mm}
\caption{Decision tree-based analysis result at location \ST 7 of \SIMPLE\ (solid edges = \true, dashed edges = \false).}
\label{fig:dt}
\end{minipage}
\end{figure*}

If we perform lifted polyhedra analysis based on the \emph{decision tree domain}
proposed in this work, then the corresponding decision tree inferred in the final program location \ST 7 of \SIMPLE\,
is depicted in Fig.~\ref{fig:dt}. 
Notice that the inner nodes of the decision tree in Fig.~\ref{fig:dt}
are labeled with \emph{Interval} linear constraints over features (\texttt{SIZE} and \texttt{B}), while
the leaves are labeled with the \emph{Polyhedra} linear constraints over program variables \texttt{x} and \texttt{y}.
Hence, we use two different numerical abstract domains in our decision trees: Interval domain \cite{DBLP:conf/popl/CousotC77}
for expressing properties in decision nodes, and Polyhedra domain \cite{DBLP:conf/popl/CousotH78} for expressing properties in leaf nodes.
The edges of decision trees are labeled with the truth value of the decision
on the parent node; we use solid edges for \true\ (i.e.\ the constraint in the parent node is satisfied) and dashed edges for \false\
(i.e.\ the negation of the constraint in the parent node is satisfied).
As decision nodes partition the space of valid configurations \Kk, we implicitly assume the correctness
of linear constraints that take into account domains of numerical features. For example, the node with constraint $(\texttt{SIZE}\!\leq\!3)$
is satisfied when $(\texttt{SIZE}\!\leq\!3) \land (1\!\leq\!\texttt{SIZE}\!\leq\!4)$, whereas its negation is satisfied when
$(\texttt{SIZE}\!>\!3) \land (1\!\leq\!\texttt{SIZE}\!\leq\!4)$.
The constraints $(1\!\leq\!\texttt{SIZE}\!\leq\!4)$ represent the domain $[1,4]$ of \texttt{SIZE}.
We can see that decision trees offer more possibilities for sharing and interaction between
analysis properties corresponding to different configurations,
they provide symbolic and compact representation of lifted analysis elements.
For example, Fig.~\ref{fig:dt} presents polyhedra properties of two program
variables \texttt{x} and \texttt{y},
which are partitioned with respect to features \texttt{B} and \texttt{SIZE}.
When ($\texttt{B} \land (\texttt{SIZE}\!\leq\!3)$) is \true\ the shared property is ($\texttt{y} \!=\!10, \texttt{x}\!=\!0$), whereas
when ($\texttt{B} \land \neg (\texttt{SIZE}\!\leq\!3)$) is \true\ the shared property is ($\texttt{y} \!=\!-10, \texttt{x}\!=\!0$).
When $\neg \texttt{B}$ is \true, the property is independent from the value of
\texttt{SIZE}, hence a node with a constraint over \texttt{SIZE}
is not needed. Therefore, all such cases are identical and so they
share the same leaf node ($\texttt{y} \!=\! 0, \texttt{x}\!=\!0$). In effect, the decision tree-based representation uses only three leafs, whereas the tuple-based representation uses eight properties.
This ability for sharing is the key motivation behind the decision trees-based representation.

\section{A Language for Program Families}

Let $\Ff = \{A_1, \ldots, A_k\}$ be a finite and totaly ordered set of \emph{numerical features} available
in a program family.
 For each feature $A \in \Ff$, $\mathrm{dom}(A) \subseteq \Zz $  denotes the set of possible values that can be assigned to $A$.
Note that any Boolean feature can be represented as a numerical feature $B \in \Ff$ with $\mathrm{dom}(B)=\{0,1\}$, such that 0 means that feature $B$
 is disabled while 1 means that $B$ is enabled.
A valid combination of feature's values represents a \emph{configuration} $k$, which specifies
one \emph{variant} of a program family.
It is given as a \emph{valuation function} $k:\Ff \to \Zz$, which is a mapping that assigns a value from $\mathrm{dom}(A)$
to each feature $A$, i.e.\ $k(A) \in \mathrm{dom}(A)$ for any $A \in \Ff$.
We assume that only a subset \(\Kk \) of all possible configurations are \emph{valid}.
An alternative representation of configurations is based upon
propositional formulae. Each configuration $k \in \Kk$ can be represented by a formula:
$(A_1=k(A_1)) \land \ldots \land (A_k=k(A_k))$.
We often abbreviate $(B=1)$ with $B$ and $(B=0)$ with $\neg B$, for a Boolean feature $B \in \Ff$.
The set of valid configurations $\Kk$ can be also represented as a formula: $\lor_{k \in \Kk} k$.

We define \emph{feature expressions}, denoted \textit{FeatExp}$(\Ff)$, as the set of propositional logic
formulas over constraints of $\Ff$ generated by the grammar:
\[
\theta ::= \true  \, | \, e_{\Ff_{\Zz}} \bowtie e_{\Ff_{\Zz}} \, | \, \neg \theta \, | \, \theta_1 \land \theta_2 \, | \, \theta_1 \lor \theta_2, \qquad
e_{\Ff_{\Zz}}::=n \mid A \mid e_{\Ff_{\Zz}} \!\oplus\! e_{\Ff_{\Zz}}
\]
where $A \in \Ff$, $n \in \Zz$, $\oplus \in \{+,-,* \}$, and $\bowtie \,\in \{=, <\}$.
We will use $\theta \in \textit{FeatExp}(\Ff)$ to write presence conditions.
When a configuration $k \in \Kk$ satisfies a feature expression $\theta \in \textit{FeatExp}(\Ff)$, we write
$k \models \theta$, where $\models$ is the standard satisfaction relation of logic.
We write $\sbr{\theta}$ to denote the set of configurations from $\Kk$ that satisfy $\theta$, that is, $k \in \sbr{\theta}$ iff $k \models \theta$.
For example, for the \SIMPLE\ program family we have $\Ff = \{\texttt{B},\texttt{SIZE}\}$, where $\mathrm{dom}(\texttt{SIZE})=[1,4]$,
and $\Kk=\{\texttt{B} \land (\texttt{SIZE}\!=\!1), \texttt{B} \land (\texttt{SIZE}\!=\!2), \texttt{B} \land (\texttt{SIZE}\!=\!3), \texttt{B} \land (\texttt{SIZE}\!=\!4),
 \neg \texttt{B} \land (\texttt{SIZE}\!=\!1), \neg \texttt{B} \land (\texttt{SIZE}\!=\!2), \neg \texttt{B} \land (\texttt{SIZE}\!=\!3), \neg \texttt{B} \land (\texttt{SIZE}\!=\!4) \}$.
 For the feature expression $(\texttt{SIZE}\!\leq\!3)$, we have
 $\sbr{(\texttt{SIZE}\!\leq\!3)}=\{\texttt{B} \land (\texttt{SIZE}\!=\!1), \texttt{B} \land (\texttt{SIZE}\!=\!2), \texttt{B} \land (\texttt{SIZE}\!=\!3),
 \neg \texttt{B} \land (\texttt{SIZE}\!=\!1), \neg \texttt{B} \land (\texttt{SIZE}\!=\!2), \neg \texttt{B} \land (\texttt{SIZE}\!=\!3) \}$.
 Hence, $B \land (\texttt{SIZE}\!=\!2) \models (\texttt{SIZE}\!\leq\!3)$ and $B \land (\texttt{SIZE}\!=\!4) \not\models (\texttt{SIZE}\!\leq\!3)$, where $B \land (\texttt{SIZE}\!=\!2) \in \Kk$, $B \land (\texttt{SIZE}\!=\!4) \in \Kk$, and $(\texttt{SIZE}\!\leq\!3) \in \textit{FeatExp}(\Ff)$.

We consider a simple sequential non-deterministic programming language,
which will be used to exemplify our work.
The program variables \Var\ are statically allocated and the only data type is the set \Zz\ of mathematical integers.
To encode multiple variants, a
new compile-time conditional statement is included.
The new statement ``$\impifdef{(\theta)}{s}$'' contains a feature
expression $\theta\in\FeatExp(\Ff)$ as a presence condition,
such that only if $\theta$ is satisfied by a configuration $k \in \Kk$
the statement $s$ will be included in the variant corresponding to $k$. The
syntax is: 
$$
\begin{array}{@{}l@{}}
s::= \impskip  \mid  \texttt{x:=}e \mid s; s \mid \texttt{if}\, (e)\, \texttt{then} \, s \, \texttt{else} \,s \mid \texttt{while}\, (e)\, \texttt{do} \, s \mid  
\texttt{\#if}\, (\theta)\, s~\texttt{\#endif}, \\
e::=n \mid [n, n'] \mid \texttt{x} \mid e \!\oplus\! e
\end{array}
$$
where $n$ ranges over integers, $[n, n']$ over integer intervals, $\texttt{x}$ over program variables \Var, and
$\oplus$ over binary arithmetic operators.
Integer intervals $[n, n']$ 
denote a random choice of an integer in the interval.
The set of all statements
$s$ is denoted by \Stm; the set of all expressions $e$ is denoted by \Exp.

A program family is evaluated in two stages. First, the C
\emph{preprocessor} \texttt{CPP} takes a program family $s$ and a configuration $k \in \Kk$ as inputs,
and produces a variant (without \texttt{\#if}-s) corresponding to $k$ as the output.
Second, the obtained variant is evaluated  using the standard single-program semantics.
The first stage is specified by the projection function $\texttt{P}_{k}$, which is an identity for all basic statements and
recursively pre-processes all sub-statements of compound statements.
Hence, $\texttt{P}_{k}(\impskip) = \impskip$
and $\texttt{P}_{k}(s ;\! s') = \texttt{P}_{k}(s) ;\! \texttt{P}_{k}(s')$.
The interesting case is ``$\impifdef{(\theta)}{s}$'', where statement $s$ is included
in the variant if $k \models \theta$, otherwise, $s$ is removed
\footnote{Since $k \in \Kk$ is a valuation function, either $k \models \theta$ holds or $k \not \models \theta$ holds for any $\theta$.}:
\[
\texttt{P}_{k}(\impifdef{(\theta)}{s}) =
    \begin{cases}
        \texttt{P}_{k}( s )  & \textrm{if} \ k \models \theta  \\
        \impskip & \textrm{if} \ k \not\models \theta
  \end{cases}
  \]
For example, variants $\texttt{P}_{B \land (\texttt{SIZE}\!=\!1)}(\!{\scriptstyle \sIMPLE}\!)$, $\texttt{P}_{B \land (\texttt{SIZE}\!=\!4)}(\!{\scriptstyle \sIMPLE}\!)$, $\texttt{P}_{\neg B \land (\texttt{SIZE}\!=\!1)}(\!{\scriptstyle \sIMPLE}\!)$, as well as $\texttt{P}_{\neg B \land (\texttt{SIZE}\!=\!4)}({\scriptstyle \sIMPLE})$ shown in
Fig.~\ref{fig:var1}, Fig.~\ref{fig:var2}, Fig.~\ref{fig:var3}, and Fig.~\ref{fig:var4}, respectively, are derived from the \SIMPLE\ family defined in Section~\ref{sec:motivate}.

\begin{figure*}[t]
\centering
\begin{subfigure}[b]{0.24\linewidth}
	$\begin{array}{@{} l @{}}
\texttt{int} \ \impassign{\texttt{x}}{10}, \, \impassign{\texttt{y}}{0}; \\
\texttt{while}\, (\texttt{x !=}\, 0)\, \, \{ \\
\qquad  \impassign{\texttt{x}}{\texttt{x-}1};  \\
\qquad  \impassign{\texttt{y}}{\texttt{y+}1};  \\
\qquad \impskip; \ \}
	\end{array}$
\vspace{-1mm}
\caption{${\scriptstyle \texttt{P}_{B \land (\texttt{SIZE}\!=\!1)}(\sIMPLE)}$}
\label{fig:var1}
\end{subfigure}
\begin{subfigure}[b]{0.24\linewidth}
$\begin{array}{@{} l @{}}
\texttt{int} \ \impassign{\texttt{x}}{10}, \, \impassign{\texttt{y}}{0}; \\
\texttt{while}\, (\texttt{x !=}\, 0)\, \, \{ \\
\qquad  \impassign{\texttt{x}}{\texttt{x-}1};  \\
\qquad  \impassign{\texttt{y}}{\texttt{y-}1};  \\
\qquad \impskip; \ \} 	
\end{array}$
\vspace{-1mm}
\caption{${\scriptstyle \texttt{P}_{B \land (\texttt{SIZE}\!=\!4)}(\sIMPLE)}$}
\label{fig:var2}
\end{subfigure}
\begin{subfigure}[b]{0.245\linewidth}
$\begin{array}{@{} l @{}}
\texttt{int} \ \impassign{\texttt{x}}{10}, \, \impassign{\texttt{y}}{0}; \\
\texttt{while}\, (\texttt{x !=}\, 0)\, \, \{ \\
\qquad  \impassign{\texttt{x}}{\texttt{x-}1};  \\
\qquad  \impassign{\texttt{y}}{\texttt{y+}1};  \\
\qquad \impassign{\texttt{y}}{0}; \ \} 	
\end{array}$
\vspace{-1mm}
\caption{${\scriptstyle \texttt{P}_{\neg \!B \land (\texttt{SIZE}\!=\!1)}(\sIMPLE)}$}
\label{fig:var3}
\end{subfigure}
\begin{subfigure}[b]{0.245\linewidth}
$\begin{array}{@{} l @{}}
\texttt{int} \ \impassign{\texttt{x}}{10}, \, \impassign{\texttt{y}}{0}; \\
\texttt{while}\, (\texttt{x !=}\, 0)\, \, \{ \\
\qquad  \impassign{\texttt{x}}{\texttt{x-}1};  \\
\qquad  \impassign{\texttt{y}}{\texttt{y-}1};  \\
\qquad \impassign{\texttt{y}}{0}; \ \} 	
\end{array}$
\vspace{-1mm}
\caption{${\scriptstyle \texttt{P}_{\neg \!B \land (\texttt{SIZE}\!=\!4)}(\sIMPLE)}$}
\label{fig:var4}
\end{subfigure}
\vspace{-4.5mm}
\caption{Different variants of the program family \SIMPLE\ from Section~\ref{sec:motivate}.} \label{fig:variants}
\end{figure*}

\section{Lifted Analysis based on Tuples}

Lifted analyses are designed by \emph{lifting} existing single-program analyses to work
on program families, rather than on individual programs.
They directly analyze program families. 
Lifted analysis as defined by Midtgaard et. al. \cite{DBLP:conf/aosd/MidtgaardBW14} rely on a lifted domain that is $|\Kk|$-fold
product of an existing single-program analysis domain \Aa\ defined over program variables \Var.
We assume that the domain $\Aa$ is equipped with sound operators
for concretization $\gamma_{\Aa}$, ordering $\sqsubseteq_{\Aa}$, join $\sqcup_{\Aa}$, meet $\sqcap_{\Aa}$, bottom $\bot_{\Aa}$,
top $\top_{\Aa}$, widening $\nabla_{\Aa}$, and narrowing $\triangle_{\Aa}$,
as well as sound transfer functions for tests $\textrm{FILTER}_{\Aa}$ and forward assignments $\textrm{ASSIGN}_{\Aa}$.
More specifically, $\textrm{FILTER}_{\Aa}(a:\Aa,e:\Exp)$ returns an abstract element from \Aa\ obtained by
restricting $a$ to satisfy the test $e$, whereas $\textrm{ASSIGN}_{\Aa}(a:\Aa,\texttt{x:=}e:\Stm)$
returns an updated version of $a$ by abstractly evaluating $\texttt{x:=}e$ in it.

\vspace{-1mm}
\paragraph*{Lifted Domain.}
The \emph{lifted analysis domain} is defined as $\langle \Aa^{\Kk}, \dot\sqsubseteq, \dot \sqcup, \dot \sqcap, \dot \bot, \dot \top \rangle$,
where $\mathbb{A}^{\Kk}$ is shorthand for
the $|\Kk|$-fold product $\prod_{k \in \Kk} \mathbb{A}$, that is, there is one separate copy of $\mathbb{A}$
for each configuration of  $\Kk$.
For example, consider the tuple in Fig.~\ref{fig:tuple}. 

\vspace{-1mm}
\paragraph*{Lifted Abstract Operations.}
Given a tuple (lifted domain element) $\overline{a} \in \Aa^{\Kk}$, the projection $\pi_{k}$ selects the ${k}^\text{th}$
component of $\overline{a}$.
All abstract lifted operations are defined by lifting
the abstract operations of the domain $\Aa$ configuration-wise.

\[
\begin{array}{ll }
\overline{\gamma}(\overline{a}) \ = \  \prod_{k \in \Kk} (\gamma_{\Aa}(\pi_{k}(\overline{a}))), \qquad &
\overline{a_1} \dot\sqsubseteq \overline{a_2} \equiv \confproj{k}{\overline{a_1}} \!\sqsubseteq_{\Aa}\!
\confproj{k}{\overline{a_2}}, \textrm{for } \forall k \!\in\! \Kk \\
\overline{a_1} ~\dot\sqcup~ \overline{a_2}   \ = \   \prod_{k \in \Kk} (\confproj{k}{\overline{a_1}} \sqcup_{\Aa} \confproj{k}{\overline{a_2}}), \ \qquad &
\overline{a_1} ~\dot\sqcap~ \overline{a_2}   \ = \   \prod_{k \in \Kk} (\confproj{k}{\overline{a_1}} \sqcap_{\Aa} \confproj{k}{\overline{a_2}}) \\
\dot \top  \ = \  \prod_{k \in \Kk} \top_{\Aa}  \ = \  (\top_{\Aa}, \ldots, \top_{\Aa}), \qquad &
\dot \bot  \ = \  \prod_{k \in \Kk} \bot_{\Aa}  \ = \  (\bot_{\Aa}, \ldots, \bot_{\Aa}) \\
\overline{a_1} ~\dot\nabla~ \overline{a_2}   \ = \  \prod_{k \in \Kk} (\confproj{k}{\overline{a_1}} \nabla_{\Aa} \confproj{k}{\overline{a_2}}), \qquad  &
\overline{a_1} ~\dot\triangle~ \overline{a_2}   \ = \   \prod_{k \in \Kk} (\confproj{k}{\overline{a_1}} \triangle_{\Aa} \confproj{k}{\overline{a_2}})
\end{array}
\]

\vspace{-1mm}
\paragraph*{Lifted Transfer Functions.}

We now define lifted transfer functions for tests, forward assignments ($\overline{\textrm{ASSIGN}}$), and \texttt{\#if}-s ($\overline{\textrm{IFDEF}}$).
There are two types of tests: \emph{expression-based tests}, denoted $\overline{\textrm{FILTER}}$, that occur in \texttt{while}-s and \texttt{if}-s, and \emph{feature-based tests}, denoted $\overline{\textrm{FEAT-FILTER}}$, that occur in \texttt{\#if}-s.
Each lifted transfer function takes as input a tuple from $\Aa^{\Kk}$ representing the invariant before evaluating
the statement (resp., expression) to handle, and returns a tuple representing the invariant after evaluating
the given statement (resp., expression).

\[
\begin{array}{l}
\overline{\textrm{FILTER}}(\overline{a}: \Aa^{\Kk}, \, e : \Exp) = {\textstyle \prod_{k \in \Kk} (\textrm{FILTER}_{\Aa}(\confproj{k}{\overline{a}},e) )} \\
\overline{\textrm{FEAT-FILTER}}(\overline{a}\!:\! \Aa^{\Kk}, \, \theta \!:\! \FeatExp(\Ff)) = {\textstyle \prod_{k \in \Kk} \begin{cases} \confproj{k}{\overline{a}},  & \textrm{if} \ k \models \theta  \\  \bot_{\Aa},  & \textrm{if} \ k \not\models \theta \end{cases} } \\
\overline{\textrm{ASSIGN}}(\overline{a}\!:\! \Aa^{\Kk}, \, \texttt{x:=}e \!:\! \Stm) = {\textstyle \prod_{k \in \Kk} (\textrm{ASSIGN}_{\Aa}(\confproj{k}{\overline{a}},\texttt{x:=}e) )} \\
\overline{\textrm{IFDEF}}(\overline{a}\!:\! \Aa^{\Kk}, \texttt{\#if} \, (\theta)\, s \!:\! \Stm) \!=\!   \overline{\sbr{s}}(\overline{\textrm{FEAT-FILTER}}(\overline{a}, \theta)) \dot\sqcup \overline{\textrm{FEAT-FILTER}}(\overline{a}, \neg \theta)
\end{array}
\]
where $\overline{\sbr{s}}(\overline{a})$ is the lifted transfer function for statement $s$.
$\overline{\textrm{FILTER}}$ and $\overline{\textrm{ASSIGN}}$ are defined by applying $\textrm{FILTER}_{\Aa}$ and
$\textrm{ASSIGN}_{\Aa}$ independently on each component of the input tuple $\overline{a}$.
$\overline{\textrm{FEAT-FILTER}}$ keeps those components $k$ of the input tuple $\overline{a}$ that satisfy $\theta$, otherwise
it replaces the other components with $\bot_{\Aa}$.
$\overline{\textrm{IFDEF}}$ captures the effect of analyzing the statement $s$ in the components $k$ of $\overline{a}$
that satisfy $\theta$, otherwise it is an identity for the other components $k$ that do not satisfy $\theta$.

\paragraph*{Lifted Analysis.}

Lifted abstract operators and transfer functions of the lifted analysis domain
$\Aa^{\Kk}$ are combined together to analyze program families.
Initially,
we build a tuple $\overline{a}_{in}$ where all components are set to $\top_{\Aa}$ for the first program location,
 and tuples where all components are set to $\bot_{\Aa}$ for all other locations.
The analysis properties are propagated forward from the first program location towards the final
location taking assignments, \texttt{\#if}-s, and tests into account with join and
widening around \texttt{while}-s.
We apply delayed widening \cite{DBLP:conf/plilp/CousotC92}, which means we start extrapolating by widening only after
some fixed number of iterations we analyze the loop.
We improve the precision of the solution obtained by delayed widening by further applying
a narrowing operator \cite{DBLP:conf/plilp/CousotC92}.
The \emph{soundness} of the lifted analysis based on $\Aa^{\Kk}$
follows immediately from the soundness of all abstract operators and transfer functions of $\Aa$ (proved in \cite{DBLP:conf/aosd/MidtgaardBW14}).

\paragraph*{Numerical Lifted Analysis}

The single-program analysis domain $\Aa$ can be instantiated by some of the well-known numerical property domains \cite{DBLP:journals/ftpl/Mine17}.

The \emph{Interval domain} \cite{DBLP:conf/popl/CousotC77}, denoted as
$\poset{\textit{I}}{\sqsubseteq_I}$, is a \emph{non-relational} numerical property domain that identifies the range of possible values for every variable as an interval.
The property elements are:
$\{ [l, h] \mid l \in \Zz \cup \{ -\infty \}, h \in \Zz \cup \{ +\infty \}, l \!\leq\! h \}$.

The \emph{Octagon domain} \cite{DBLP:journals/lisp/Mine06}, denoted as
$\poset{\textit{O}}{\sqsubseteq_O}$, is a \emph{weakly-relational} numerical property domain, where property elements are
 conjunctions of linear constraints of the form $\mypm \texttt{x}_j \mypm \texttt{x}_i \geq \beta$
between variables \texttt{x}$_i$ and \texttt{x}$_j$, and $\beta \in \Zz$. 

The \emph{Polyhedra domain} \cite{DBLP:conf/popl/CousotH78}, denoted as
$\poset{\textit{P}}{\sqsubseteq_P}$, is a \emph{fully relational} numerical property domain.
It expresses conjunctions of linear constraints of the form $\alpha_1 \texttt{x}_1 + \ldots + \alpha_n \texttt{x}_n + \beta \geq 0$, where
 \texttt{x}$_1$, $\ldots$, \texttt{x}$_n$ are variables and $\alpha_i, \beta \in \Zz$.

\section{Lifted Analysis based on Decision Trees} \label{sec:analysis}

We now introduce a new \emph{decision tree }lifted domain. 
Its elements 
are disjunctions of leaf nodes that
belong to an existing single-program domain $\Aa$ defined over program variables \Var.
The leaf nodes are separated
by linear constraints over numerical features, organized in the decision nodes.
Hence, we encapsulate the set of configurations $\Kk$ into the decision nodes
of a decision tree where each top-down path represents one or several configurations that satisfy the
constraints encountered along the given path. We store
in each leaf node the property generated from the variants representing the corresponding configurations.


\paragraph*{Abstract domain for decision nodes.}
We define the family of abstract domains for linear constraints $\Cc_{\Dd}$,
which are parameterized by any of the numerical property domains \Dd\, (intervals $\textit{I}$, octagons $\textit{O}$, polyhedra $\textit{P}$).
We use $C_{\textit{I}} = \{ \mypm A_i \geq \beta \mid A_i \in \Ff, \beta \in \Zz \}$ to denote the set of \emph{interval constraints},
$C_{\textit{O}} = \{ \mypm A_i \mypm A_j \geq \beta \mid A_i, A_j \in \Ff, \beta \in \Zz \}$ to denote the set of \emph{octagonal constraints},
and $C_{\textit{P}} = \{ \alpha_1 A_1 + \ldots + \alpha_k A_k + \beta \geq 0 \mid A_1, \ldots A_k \in \Ff, \alpha_1, \ldots, \alpha_k, \beta \in \Zz, \textrm{gcd}(|\alpha_1|, \ldots, |\alpha_k|,|\beta|)=1 \}$ to denote the set of \emph{polyhedral constraints}. We have $C_{\textit{I}} \subseteq C_{\textit{O}} \subseteq C_{\textit{P}}$.

The set $C_{\Dd}$ of linear constraints over features $\Ff$ is constructed by the underlying numerical property domain
$\poset{\Dd}{\sqsubseteq_{\Dd}}$ using the Galois connection $\poset{\mathcal{P}(C_{\Dd})}{\sqsubseteq_{\Dd}} \galois{\alpha_{C_{\Dd}}}{\gamma_{C_{\Dd}}} \poset{\Dd}{\sqsubseteq_{\Dd}}$, where $\mathcal{P}(C_{\Dd})$ is the power
set of $C_{\Dd}$.
The abstraction function $\alpha_{C_{\Dd}}: \mathcal{P}(C_{\Dd}) \to \Dd$ maps a set of interval (resp., octagon, polyhedral) constraints
to an interval (resp., an octagon, polyhedral) that represents a conjunction of constraints; the concretization function
$\gamma_{C_{\Dd}}: \Dd \to \mathcal{P}(C_{\Dd})$ maps an interval (resp., an octagon, a polyhedron) that represents a conjunction of constraints to a set of interval
(resp., octagonal, polyhedral) constraints. We have $\gamma_{C_{\Dd}}(\top_{\Dd}) = \emptyset$ and $\gamma_{C_{\Dd}}(\bot_{\Dd}) = \{ \bot_{C_{\Dd}} \}$, where $\bot_{C_{\Dd}}$ is an unsatisfiable constraint. 

The domain of decision nodes is $\Cc_{\Dd}$. We assume $\Ff = \{A_1, \ldots, A_k\}$ be a finite and totally ordered set of
features, such that the ordering is $A_1 > A_2 > \ldots > A_{k}$.
We impose a total order $<_{\Cc_{\Dd}}$ on $\Cc_{\Dd}$ to be the lexicographic order on the coefficients $\alpha_1, \ldots, \alpha_{k}$ and constant $\alpha_{k+1}$ of the linear constraints, such that:
\[
\begin{array}{l}
(\alpha_1 \cdot A_1 + \ldots + \alpha_{k} \cdot A_{k} + \alpha_{k+1} \!\geq\! 0) ~<_{\Cc_{\Dd}}~
(\alpha'_1 \cdot A_1 + \ldots + \alpha'_{k} \cdot A_{k} + \alpha'_{k+1} \!\geq\! 0) \\
~\iff~ \exists j >0. \forall i<j. (\alpha_i = \alpha'_i) \land (\alpha_j < \alpha'_j)
\end{array}
\]

The negation of linear constraints is formed as:
$\neg (\alpha_1 A_{1} + \ldots \alpha_k A_{k} + \beta \geq 0) = -\alpha_1 A_{1} - \ldots - \alpha_k A_{k} -\beta - 1 \geq 0$.
For example, the negation of $A - 3 \geq 0$ is the constraint $-A + 2 \geq 0$ (i.e., $A \leq 2$).
To ensure canonical representation of decision trees, a linear constraint $c$ and its negation $\neg c$
cannot both appear as nodes in a decision tree.
For example, we only keep the largest constraint with respect to $<_{C_{\Dd}}$ between $c$ and $\neg c$.
For this reason, we define the equivalence relation $\equiv_{C_{\Dd}}$ as $c \equiv_{C_{\Dd}} \neg c$.
We define $\poset{\Cc_{\Dd}}{<_{\Cc_{\Dd}}}$ to denote $\poset{C_{\Dd}/_{\equiv}}{<_{C_{\Dd}}}$,
such that elements of $\Cc_{\Dd}$ are constraints obtained by quotienting by the equivalence $\equiv_{C_{\Dd}}$.

\paragraph*{Abstract domain for constraint-based decision trees.}

A \emph{constraint-based decision tree} $t \in \Tt(\Cc_{\Dd}, \Aa)$ over
the sets $\Cc_{\Dd}$ of linear constraints defined over $\Ff$ and the leaf abstract domain $\Aa$ defined over \Var\ is either
a leaf node $\san{a}$ with $a \in \Aa$, or
$\sbr{c:tl,tr}$, where $c \in \Cc_{\Dd}$ (denoted by $t.c$) is the smallest constraint with respect to
$<_{\Cc_{\Dd}}$ appearing in the tree $t$, $tl$ (denoted by $t.l$) is the left
subtree of $t$ representing its \emph{\true\, branch}, and $tr$ (denoted by $t.r$) is the right
subtree of $t$ representing its \emph{\false\, branch}. 
The path along a decision tree establishes the set of configurations (those that satisfy the encountered constraints),
and the leaf nodes represent the analysis properties for the corresponding configurations.

\begin{example} \label{exp:dt1}
The following two constraint-based decision trees $t_1$ and $t_2$ have decision nodes labelled with Interval linear constraints
over the numeric feature \texttt{SIZE} with domain $\{1, 2, 3, 4\}$, whereas leaf nodes are Interval properties:
\[
\begin{array}{l}
t_1 = \sbr{\texttt{SIZE} \!\geq\! 4: \san{[y \!\geq\! 2]}, \san{[y \!=\! 0]}}, \
t_2 = \sbr{\texttt{SIZE} \!\geq\! 2: \san{[y \!\geq\! 0]}, \san{[y \!\leq\! 0]}} \  \qed
\end{array}
\]
\end{example}

\paragraph*{Abstract Operations.}
The \emph{concretization function} $\gamma_{\Tt}$ of a decision tree $t \in \Tt(\Cc_{\Dd}, \Aa)$
returns $\gamma_{\Aa}(a)$ for $k \in \Kk$, where $k$ satisfies the set $C \in \mathcal{P}(\Cc_{\Dd})$
of constraints accumulated along the top-down path to the leaf node $a \in \Aa$.
More formally, $\gamma_{\Tt}(t) = \overline{\gamma}_{\Tt}[\Kk](t)$.
The function $\overline{\gamma}_{\Tt}$ accumulates into a set $C \in \mathcal{P}(\Cc_{\Dd})$
constraints along the paths up to a leaf node, which is
initially equal to the set of implicit constraints over $\Ff$, $\Kk \!=\! \lor_{k \in \Kk} k$,
taking into account domains of features:
\[
\overline{\gamma}_{\Tt}[C](\san{a}) \!=\! {\textstyle \prod_{k \models C}}\gamma_{\Aa}(a), \quad \overline{\gamma}_{\Tt}[C](\sbr{c\!:\!tl,tr}) \!=\! \overline{\gamma}_{\Tt}[C \cup \{c\}](tl) \times  \overline{\gamma}_{\Tt}[C \cup \{\neg c\}](tr)
\]
Note that $k \models C$ is equivalent with $\alpha_{\Cc_{\Dd}}(\{k\}) \sqsubseteq_{\Dd} \alpha_{\Cc_{\Dd}}(C)$. 
Therefore, we can check $k \models C$ using the abstract operation $\sqsubseteq_{\Dd}$ of the numerical domain \Dd.

Other binary operations of  $\Tt(\Cc_{\Dd}, \Aa)$ are based on Algorithm~\ref{Algorithm1} for \emph{tree unification},
which finds a common refinement (labelling) of two trees $t_1$ and $t_2$ by calling function $\texttt{UNIFICATION}(t_1,t_2,\Kk)$.
It possibly adds new constraints as decision nodes (Lines 5--7, Lines 11--13), or removes constraints that are redundant (Lines 3,4,9,10,15,16).
The function \texttt{UNIFICATION} accumulates into the set $C \in \mathcal{P}(\Cc_{\Dd})$ (initialized to $\Kk$, which represents implicit constraints satisfied by both $t_1$ and $t_2$),
constraints encountered along the paths of the decision tree. This set $C$ is used by
the function $\mathrm{isRedundant}(c,C)$, which checks whether the linear constraint $c \in \Cc_{\Dd}$ is redundant with respect
to $C$ by testing $\alpha_{\Cc_{\Dd}}(C) \sqsubseteq_{\Dd} \alpha_{\Cc_{\Dd}}(\{c\})$.
Note that the tree unification does not lose any information.

\begin{algorithm}
\lIf{$\mathrm{isLeaf}(t_1) \land \mathrm{isLeaf}(t_2)$}{\bf{return} ($t_1,t_2$)}
\If{$\mathrm{isLeaf}(t_1) \lor (\mathrm{isNode}(t_1) \land \mathrm{isNode}(t_2) \land t_2.c <_{\Cc_{\Dd}} t_1.c)$}  {
\lIf{$\mathrm{isRedundant}(t_2.c,C)$}{\bf{return} \texttt{UNIFICATION($t_1,t_2.l,C$)}}
\lIf{$\mathrm{isRedundant}(\neg t_2.c,C)$}{\bf{return} \texttt{UNIFICATION($t_1,t_2.r,C$)}}
$(l_1,l_2) =$ \texttt{UNIFICATION($t_1,t_2.l,C \cup \{t_2.c\}$)}\;
$(r_1,r_2) =$ \texttt{UNIFICATION($t_1,t_2.r,C \cup \{\neg t_2.c\}$)}\;
\bf{return} ($\sbr{t_2.c:l_1,r_1},\sbr{t_2.c:l_2,r_2}$)\;
}
\eIf{$\mathrm{isLeaf}(t_2) \lor (\mathrm{isNode}(t_1) \land \mathrm{isNode}(t_2) \land t_1.c <_{\Cc_{\Dd}} t_2.c)$}  {
\lIf{$\mathrm{isRedundant}(t_1.c,C)$}{\bf{return} \texttt{UNIFICATION($t_1.l,t_2,C$)}}
\lIf{$\mathrm{isRedundant}(\neg t_1.c,C)$}{\bf{return} \texttt{UNIFICATION($t_1.r,t_2,C$)}}
$(l_1,l_2) =$ \texttt{UNIFICATION($t_1.l,t_2,C \cup \{t_1.c\}$)}\;
$(r_1,r_2) =$ \texttt{UNIFICATION($t_1.r,t_2,C \cup \{\neg t_1.c\}$)}\;
\bf{return} ($\sbr{t_1.c:l_1,r_1},\sbr{t_1.c:l_2,r_2}$)\;
}
{
\lIf{$\mathrm{isRedundant}(t_1.c,C)$}{\bf{return} \texttt{UNIFICATION($t_1.l,t_2.l,C$)}}
\lIf{$\mathrm{isRedundant}(\neg t_1.c,C)$}{\bf{return} \texttt{UNIFICATION($t_1.r,t_2.r,C$)}}
$(l_1,l_2) =$ \texttt{UNIFICATION($t_1.l,t_2.l,C \cup \{t_1.c\}$)}\;
$(r_1,r_2) =$ \texttt{UNIFICATION($t_1.r,t_2.r,C \cup \{\neg t_1.c\}$)}\;
\bf{return} ($\sbr{t_1.c:l_1,r_1},\sbr{t_1.c:l_2,r_2}$)\; } 
\caption{{\bf \texttt{UNIFICATION($t_1,t_2,C$)}} \label{Algorithm1}}
\end{algorithm}

\begin{example} \label{exp:dt2}
Consider constraint-based decision trees $t_1$ and $t_2$ from Example~\ref{exp:dt1}.  After tree unification
$\texttt{UNIFICATION}(t_1,t_2,\Kk)$, the resulting decision trees are:
\[
\begin{array}{l}
t_1 = \sbr{\texttt{SIZE} \geq 4: \san{[y \geq 2]}, \sbr{\texttt{SIZE} \geq 2: \san{[y = 0]}, \san{[y = 0]}}}, \\
t_2 = \sbr{\texttt{SIZE} \geq 4: \san{[y \geq 0]}, \sbr{\texttt{SIZE} \geq 2: \san{[y \geq 0]}, \san{[y \leq 0]}}}
\end{array}
\]
Note that \texttt{UNIFICATION} adds a decision node for $\texttt{SIZE} \geq 2$ to the right subtree of $t_1$,
whereas it adds a decision node for $\texttt{SIZE} \geq 4$ to $t_2$ and removes the redundant constraint
$\texttt{SIZE} \geq 2$ from the resulting left subtree of $t_2$. \qed
\end{example}

All binary operations are performed leaf-wise on the unified decision trees.
Given two unified decision trees $t_1$ and $t_2$, their ordering 
and join 
are  defined as:
\[
\begin{array}{l}
\san{a_1} \, \sqsubseteq_{\Tt}\,  \san{a_2} = a_1 \!\sqsubseteq_{\Aa}\! a_2, \!\!\quad
\sbr{c\!:\!tl_1,tr_1} \!\sqsubseteq_{\Tt}\! \sbr{c\!:\!tl_2,tr_2} \!=\! (tl_1 \!\sqsubseteq_{\Tt}\! tl_2) \land (tr_1 \!\sqsubseteq_{\Tt}\! tr_2) \\
\san{a_1} \!\sqcup_{\Tt}\! \san{a_2} = \san{a_1 \!\sqcup_{\Aa}\! a_2}, \!\quad
\sbr{c\!:\!tl_1,tr_1} \!\sqcup_{\Tt}\! \sbr{c\!:\!tl_2,tr_2} \!=\! \sbr{c:tl_1 \!\sqcup_{\Tt}\! tl_2, tr_1 \!\sqcup_{\Tt}\! tr_2}
\end{array}
\]
Similarly, we compute meet, 
widening, 
and narrowing 
of 
$t_1$ and $t_2$.
The top is a tree with a single $\top_{\Aa}$ leaf: $\top_{\Tt} = \san{\top_{\Aa}}$, while the bottom is: 
$\bot_{\Tt} = \san{\bot_{\Aa}}$.

\begin{example} \label{exp:dt3}
Consider the unified trees $t_1$ and $t_2$ from Example~\ref{exp:dt2}.  We have that
$t_1 \!\sqsubseteq_{\Tt}\! t_2$ holds, and
$t_1 \!\sqcup_{\Tt}\! t_2 \!=\! \sbr{\texttt{SIZE} \!\geq\!\! 4\!: \san{[y \!\geq\! 0]}, \sbr{\texttt{SIZE} \!\geq\!\! 2\!: \san{[y \!\geq\! 0]}, \san{[y \!\leq\! 0]}}}$. 
\end{example}

\paragraph*{Transfer functions.}
The transfer functions for forward assignments ($\textrm{ASSIGN}_{\Tt}$) and expression-based tests ($\textrm{FILTER}_{\Tt}$)
modify only leaf nodes of a constraint-based decision tree.
In contrast, transfer functions for variability-specific constructs, such as feature-based tests ($\textrm{FEAT-FILTER}_{\Tt}$) and \texttt{\#if}-s  ($\textrm{IFDEF}_{\Tt}$) add, modify, or delete decision nodes of a decision tree.
This is due to the fact that the analysis information about program variables is located in leaf nodes, while
the information about feature variables is located in decision nodes.

\begin{algorithm}
\lIf{$\mathrm{isLeaf}(t)$}{\bf{return} $\san{\texttt{ASSIGN}_{\Aa}(t,\texttt{x:=}e)}$}
\bf{return} $\sbr{t.c:\texttt{ASSIGN}_{\Tt}(t.l,\texttt{x:=}e$),\texttt{ASSIGN}$_{\Tt}(t.r,\texttt{x:=}e)}$\;
\caption{{\bf \texttt{ASSIGN}$_{\Tt}(t,\texttt{x:=}e$)} \label{Algorithm3}}
\end{algorithm}

Transfer function $\textrm{ASSIGN}_{\Tt}$ for handling  an assignment $\texttt{x:=}e$ in the input tree $t$
is described by Algorithm~\ref{Algorithm3}.
Note that $\texttt{x}\in \Var$, and $e \in \Exp$ may contain only program variables.
 We apply $\textrm{ASSIGN}_{\Aa}$ to each leaf node $a$ of $t$, which substitutes expression $e$ for variable \texttt{x} in $a$.
Similarly, transfer function $\textrm{FILTER}_{\Tt}$ for handling expression-based tests $e \in Exp$ is implemented
by applying $\textrm{FILTER}_{\Aa}$ leaf-wise. 

Transfer function $\textrm{FEAT-FILTER}_{\Tt}$ for feature-based tests $\theta$ is
described by Algorithm~\ref{Algorithm5}.
It reasons by induction on the structure of $\theta$ (we assume negation is applied to atomic propositions).
When $\theta$ is an atomic constraint over numerical features (Lines 2,3), we use $\textrm{FILTER}_{\Dd}$
to approximate $\theta$, thus
producing a set of constraints $J$, which are then added to the tree $t$, possibly
discarding all paths of $t$ that do not satisfy $\theta$. This is done by calling function $\texttt{RESTRICT}(t,\Kk,J)$,
which adds linear constraints from $J$ to $t$ in ascending order with respect to $<_{\Cc_{\Dd}}$ as shown in Algorithm~\ref{Algorithm:restrict}. 
Note that $\theta$ may not be representable exactly in $\Cc_{\Dd}$ (e.g., in the case of non-linear constraints over $\Ff$),
so $\textrm{FILTER}_{\Dd}$ may produce a set of constraints approximating it.
When $\theta$ is a conjunction (resp., disjunction) of two
feature expressions (Lines 4,5) (resp., (Lines 6,7)), the resulting decision trees are
merged by operation meet $\sqcap_{\Tt}$ (resp., join $\sqcup_{\Tt}$).
Function $\texttt{RESTRICT}(t,C,J)$, described in Algorithm~\ref{Algorithm:restrict}, takes as input a decision tree $t$, a set $C$ of linear constraints
accumulated along paths up to a node, and a set $J$ of linear constraints in canonical form that need to be added to $t$.
For each constraint $j \in J$, there exists a boolean $b_j$ that shows whether the tree should be
constrained with respect to $j$ or with respect to $\neg j$.
When $J$ is not empty, the linear constraints from $J$ are added to $t$ in ascending order with respect to $<_{\Cc_{\Dd}}$.
At each iteration, the smallest linear constraint $j$ is extracted from $J$ (Line 9), and is handled
appropriately based on whether $j$ is smaller (Line 11--15), or greater or equal (Line 17--21) to the
constraint at the node of $t$ we currently consider.

\begin{algorithm}[t]
\DontPrintSemicolon
\SetKwProg{Fn}{Function}{:}{}
\Switch{$\theta$}{
  \Case{$(e_{\Ff_{\Zz}} \bowtie e_{\Ff_{\Zz}}) ~||~ (\neg (e_{\Ff_{\Zz}} \bowtie e_{\Ff_{\Zz}}))$ }
  {$J = \texttt{FILTER}_{\Dd}(\top_{\Dd},\theta); \ $\bf{return} $\texttt{RESTRICT}(t,\Kk,J)$\;}
  \Case{$\theta_1 \land \theta_2$} {\bf{return} $\texttt{FEAT-FILTER}_{\Tt}(t,\theta_1) \sqcap_{\Tt} \texttt{FEAT-FILTER}_{\Tt}(t,\theta_2)$}
  \Case{$\theta_1 \lor \theta_2$} {\bf{return} $\texttt{FEAT-FILTER}_{\Tt}(t,\theta_1) \sqcup_{\Tt} \texttt{FEAT-FILTER}_{\Tt}(t,\theta_2)$}
}
\caption{{\bf \texttt{FEAT-FILTER}$_{\Tt}(t,\theta$)} \label{Algorithm5}}
\end{algorithm}

\begin{algorithm}
\eIf{$\mathrm{isEmpty}(J)$}
{
\lIf{$\mathrm{isLeaf}(t)$}{return $t$}
\lIf{$\mathrm{isRedundant}(t.c,C)$}{return \texttt{RESTRICT}($t.l,C,J$)}
\lIf{$\mathrm{isRedundant}(\neg t.c,C)$}{return \texttt{RESTRICT}($t.r,C,J$)}
$l = \texttt{RESTRICT}(t.l,C \cup \{t.c\},J)$ \;
$r = \texttt{RESTRICT}(t.r,C \cup \{\neg t.c\},J)$ \;
\bf{return} ($\sbr{t.c:l,r}$)\;
 } {
$j = min_{<_{\Cc_{\Dd}}}(J)$ \;
\eIf{$\mathrm{isLeaf}(t) \lor (\mathrm{isNode}(t) \land j <_{\Cc_{\Dd}} t.c)$}  {
\lIf{$\mathrm{isRedundant}(j,C)$}{return \texttt{RESTRICT($t,C,J\backslash\{j\}$)}}
\lIf{$\mathrm{isRedundant}(\neg j,C)$}{return $\san{\bot_{\Aa}}$}
\lIf{$j =_{\Cc_{\Dd}} t.c$}{(\bf{if} $b_j$ \bf{then} $t=t.l$; \bf{else} $t=t.r$) }
\lIf{$b_j$}{\bf{return} ($\sbr{j:\texttt{RESTRICT}(t,C\cup\{j\},J\backslash\{j\}),\san{\bot_{\Aa}}}$) }
\lElse{\bf{return} ($\sbr{j:\san{\bot_{\Aa}},\texttt{RESTRICT}(t,C\cup\{j\},J\backslash\{j\})}$) }
}{
\lIf{$\mathrm{isRedundant}(t.c,C)$}{return \texttt{RESTRICT}($t.l,C,J$)}
\lIf{$\mathrm{isRedundant}(\neg t.c,C)$}{return \texttt{RESTRICT}($t.r,C,J$)}
$l = \texttt{RESTRICT}(t.l,C \cup \{t.c\},J)$ \;
$r = \texttt{RESTRICT}(t.r,C \cup \{\neg t.c\},J)$ \;
\bf{return} ($\sbr{t.c:l,r}$)\;
} }
\caption{{\bf \texttt{RESTRICT($t,C,J$)}} \label{Algorithm:restrict}}
\end{algorithm}

Finally, transfer function \textrm{IFDEF}$_{\Tt}$ is defined as:
\[
\textrm{IFDEF}_{\Tt}(t, \, \texttt{\#if} \, (\theta)\, s) =  \sbr{s}_{\Tt}(\textrm{FEAT-FILTER}_{\Tt}(t, \, \theta)) \, \sqcup_{\Tt} \, \textrm{FEAT-FILTER}_{\Tt}(t, \, \neg \theta)
\]
where $\sbr{s}_{\Tt}(t)$ denotes the transfer function in $\Tt(\Cc_{\Dd}, \Aa)$ for statement $s$.

After applying transfer functions,
the obtained decision trees may contain some redundancy that can be exploited to further compress them.
Function $\texttt{COMPRESS}_{\Tt}(t,C)$, described by Algorithm~\ref{Algorithm6}, is applied to decision trees $t$
in order to compress (reduce) their representation. We use five different optimizations.
First, if constraints on a path to some leaf are unsatisfiable, we eliminate that leaf node (Lines 9,10).
Second, 
if a decision node contains two same subtrees, then we keep only one subtree and we also eliminate the decision node (Lines 11--13).
Third, if a decision node contains a left leaf and a right subtree, such that its left leaf is the same with the left leaf of its right subtree
and the constraint in the decision node is less or equal to the constraint in the root of its right subtree, then
we can eliminate the decision node and its left leaf (Lines 14,15).
A similar rule exists when a decision node has a left subtree and a right leaf (Lines 16,17).


\begin{algorithm}[t]
\Switch{$t$}{
  \Case{$\san{n}$ }
  {\bf{return} $\san{n}$\;}
  \Case{$\sbr{t.c:l,r}$} {
  $l' = \texttt{COMPRESS}_{\Tt}(t.l,C \cup \{t.c\})$ \;
  $r' = \texttt{COMPRESS}_{\Tt}(t.r,C \cup \{\neg t.c\})$ \;
  \Switch{$l',r'$}{
          \Case{$\san{n'_{l}},\san{n'_{r}}$} {
            \lIf{$\mathrm{UNSAT}(C \cup \{t.c\})$}{return $\san{n'_{r}}$}
            \lIf{$\mathrm{UNSAT}(C \cup \{\neg t.c\})$}{return $\san{n'_{l}}$}
            \lIf{$n'_{l}=n'_{r}$}{return $\san{n'_{l}}$}}
          \Case{$\sbr{c_1:l_1,r_1},\sbr{c_2:l_2,r_2}$ when $c_1=c_2 \land l_1=l_2 \land r_1=r_2$} {
            \bf{return} $\sbr{c_1:l_1,r_1}$\;}
          \Case{$\san{n'_{l}},\sbr{c_2:l_2,r_2}$ when $\san{n'_{l}} = l_2 \land c \leq_{\Cc_{\Dd}} c_2$} {
            \bf{return} $\sbr{c_2:l_2,r_2}$\;}
          \Case{$\sbr{c_1:l_1,r_1},\san{n'_{r}}$ when $\san{n'_{r}} = r_1 \land c_1 \leq_{\Cc_{\Dd}} c$} {
            \bf{return} $\sbr{c_1:l_1,r_1}$\;}
          \Case{default:} {
            \bf{return} $\sbr{t.c:l',r'}$\;}
    }
    }
    }
\caption{{\bf \texttt{COMPRESS}$_{\Tt}(t,C$)} \label{Algorithm6}}
\end{algorithm}

\paragraph*{Lifted analysis.}
The abstract operations and transfer functions of $\Tt(\Cc_{\Dd}, \Aa)$ can be used to define the
lifted analysis for program families.
Tree $t_{in}$ at the initial location has only one leaf node
$\top_{\Aa}$ and decision nodes that define the set \Kk.
Note that if $\Kk \equiv \true$, then $t_{in}=\top_{\Tt}$.
In this way, we collect the possible invariants in the form of decision trees at all program locations.

We establish correctness of the lifted analysis based on $\Tt(\Cc_{\Dd}, \Aa)$ by showing that it produces
identical results with tuple-based domain $\Aa^{\Kk}$.
Let $\sbr{s}_{\Tt}$ and $\overline{\sbr{s}}$ denote transfer functions of statement $s$ in
$\Tt(\Cc_{\Dd}, \Aa)$ and $\Aa^{\Kk}$, respectively.
\begin{theorem}[App.~\ref{app:alg}] \label{thm:1}
$\gamma_{\Tt}\big( \sbr{s}_{\Tt}(t_{in}) \big) = \overline{\gamma} \big( \overline{\sbr{s}}(\overline{a}_{in})  \big)$.
\end{theorem}

\begin{example}\label{exp:dt4}
Let us consider the code base of a program family $P$ given in Fig.~\ref{fig:dt4:code}.
It contains only one numerical feature \texttt{SIZE} with domain $\Nn$.
The decision tree inferred at the final location \ST 4 is depicted in Fig.~\ref{fig:dt4:dt}.
It uses the Interval domain for both decision and leaf nodes.
Note that the constraint $(\texttt{SIZE} \!<\! 3)$ does not explicitly appear in the code base, but we
obtain it in the decision tree representation. 
This shows that partitioning of the configuration space \Kk\, induced by decision trees
is semantics-based rather than syntactic-based.
\end{example}
\begin{figure*}[t]
\centering
\begin{minipage}[b]{0.54\linewidth}
\begin{center}
	$\begin{array}{ll}
\textcolor{gray}{{\ST 1}} \  & \Scale[0.89]{\texttt{int} \ \impassign{\texttt{x}}{0};} \\
\textcolor{gray}{{\ST 2}} \  & \Scale[0.89]{\texttt{\#if} \, (\texttt{SIZE} \leq 4) \ \impassign{\texttt{x}\!}{\!\texttt{x+}1}; \ \texttt{\#else} \ \impassign{\texttt{x\!}}{\!\texttt{x-}1}; \ \texttt{\#endif}} \\
\textcolor{gray}{{\ST 3}} \  & \Scale[0.89]{\texttt{\#if} \, (\texttt{SIZE==3} \, || \, \texttt{SIZE==4}) \ \impassign{\texttt{x}}{\texttt{x-}2}; \ \texttt{\#endif} \ \textcolor{gray}{{\ST 4}}}
	\end{array}$
\end{center}
\caption{Code base for program family $P$.}
\label{fig:dt4:code}
\end{minipage}
\begin{minipage}[b]{0.44\linewidth}
\centering
\begin{tikzpicture}[-,>=stealth',level/.style={sibling distance = 1.25cm/#1,
  level distance = 0.4cm}]
     \node[arn_n](A){\tiny{$\texttt{SIZE}\!\!<\!\!\!3$}};
     \node[arn_x,below left=of A](B){${\scriptstyle [\texttt{x=} 1]}$};
     \node[arn_x,below right=of A](C){${\scriptstyle [\texttt{x=-} 1]}$};
     \draw[dashed](A) to node[above right]{} (C);
     \draw[-](A) to node[above left]{} (B);
  \end{tikzpicture}
\caption{Decision tree at loc. \!\ST 4 of $P$.}
\label{fig:dt4:dt}
\end{minipage}
\end{figure*}

\begin{example}\label{exp:dt5}
Let us consider the code base of a program family $P'$ given in Fig.~\ref{fig:dt5:code}.
It contains one numerical feature \texttt{A} with domain $[1,4]$ and a non-linear feature expression $\texttt{A}*\texttt{A}<9$.
At program location \ST 2, \texttt{FEAT-FILTER}$_{\Tt}(\san{\texttt{x}=0},\texttt{A}*\texttt{A}<9$) returns an
over-approximating tree $\san{\texttt{x}=0}$, whereas \texttt{FEAT-FILTER}$_{\Tt}(\san{\texttt{x}=0},\neg (\texttt{A}*\texttt{A}<9)$) returns  $\sbr{\texttt{A}\!\geq\!3,\san{\texttt{x}=0},\san{\bot_{I}}}$. In effect, we obtain
an over-approximating result
at the final program location \ST 3 as shown in Fig.~\ref{fig:dt5:dt1}.
The precise result at the program location \ST 3, which can be obtained in case we have numerical domains that can handle non-linear constraints,
is given in Fig.~\ref{fig:dt5:dt2}.
We observe that when $\neg (\texttt{A} \leq 2)$, we obtain an over-approximating analysis result ($-1 \!\leq\! \texttt{x} \!\leq\! 1$ instead of $\texttt{x}=-1$) due to the
over-approximation of the non-linear feature expression in the numerical domains we use. \qed
\end{example}
\begin{figure*}[t]
\centering
\begin{minipage}[b]{0.32\linewidth}
\begin{center}
	$\begin{array}{ll}
\textcolor{gray}{{\ST 1}} & \Scale[0.89]{\texttt{int} \ \impassign{\texttt{x}}{0};} \\
\textcolor{gray}{{\ST 2}} & \Scale[0.89]{\texttt{\#if} \, (\texttt{A}*\texttt{A}<9) \ \impassign{\texttt{x}}{\texttt{x+}1};} \\
& \Scale[0.89]{ \texttt{\#else} \ \impassign{\texttt{x}}{\texttt{x-}1}; \ \texttt{\#endif}} \, \textcolor{gray}{{\ST 3}}
	\end{array}$
\end{center}
\caption{Code base for $P'$.}
\label{fig:dt5:code}
\end{minipage}
\begin{minipage}[b]{0.33\linewidth}
\centering
\begin{tikzpicture}[-,>=stealth',level/.style={sibling distance = 1cm/#1,
  level distance = 0.3cm}]
     \node[arn_n](A){\tiny{$\texttt{A}\!\!\leq\!\!2$}};
     \node[arn_x,below left=of A](B){${\scriptstyle [\texttt{x=} 1]}$};
     \node[arn_x,below right=of A](C){${\scriptstyle [-1 \!\leq\! \texttt{x} \!\leq\! 1]}$};
     \draw[dashed](A) to node[above right]{} (C);
     \draw[-](A) to node[above left]{} (B);
  \end{tikzpicture}
\caption{Over-approximating decis. tree at loc. \ST 3 of $P'$.}
\label{fig:dt5:dt1}
\end{minipage}
\begin{minipage}[b]{0.33\linewidth}
\centering
\begin{tikzpicture}[-,>=stealth',level/.style={sibling distance = 1cm/#1,
  level distance = 0.3cm}]
     \node[arn_n](A){\tiny{$\texttt{A}\!\!\leq\!\!2$}};
     \node[arn_x,below left=of A](B){${\scriptstyle [\texttt{x=} 1]}$};
     \node[arn_x,below right=of A](C){${\scriptstyle [\texttt{x=-} 1]}$};
     \draw[dashed](A) to node[above right]{} (C);
     \draw[-](A) to node[above left]{} (B);
  \end{tikzpicture}
\caption{Precise decision tree at loc. \ST 3 of $P'$.}
\label{fig:dt5:dt2}
\end{minipage}
\vspace{-1mm}
\end{figure*}

\vspace{-2mm}
\section{Evaluation} \label{sec:evaluation}


\paragraph*{Implementation}
We have developed a prototype lifted static analyzer, called \SPL, that uses lifted abstract domains of tuples $\Aa^{\Kk}$ and decision trees
$\Tt(\Cc_{\Dd}, \Aa)$.
The abstract domains $\Aa$ for encoding properties of tuple components and leaf nodes
as well as the abstract domain $\Dd$ for encoding linear constraints over numerical features
are based
on intervals, octagons, and polyhedra domains. Their
abstract operations and transfer functions are provided by the \APRON\, library \cite{DBLP:conf/cav/JeannetM09}.
Our proof-of-concept implementation is written in \textsc{OCaml} and consists of around 6K lines of code.
The current front-end of the tool accepts programs written in
a (subset of) \textsc{C} with \texttt{\#if} directives, but without \texttt{struct} and \texttt{union} types.
It currently provides only a limited support for arrays, pointers, and recursion. 
The only basic data type is mathematical integers.
\SPL\ automatically infers numerical invariants in all program locations corresponding to all variants in the given family.

\paragraph*{Experimental setup and Benchmarks}
All experiments are executed on a 64-bit Intel$^\circledR$Core$^{TM}$
i7-8700 CPU@3.20GHz $\times$ 12, Ubuntu 18.04.5 LTS, with 8 GB memory, and we use a timeout value of 300 sec.
All times are reported as average over five independent executions.
The implementation, benchmarks, and all results obtained from our experiments
are available from: \url{http://bit.ly/2SRElgK}. 
In our experiments, we use three instances of our lifted analysis via decision trees:
$\overline{\mathcal A}_{\Tt}(I)$, $\overline{\mathcal A}_{\Tt}(O)$, and $\overline{\mathcal A}_{\Tt}(P)$ that use
 intervals, octagons, and polyhedra domains for properties in leaf nodes and in decision nodes, respectively.
We also use three instances of our lifted analysis based on tuples:
$\overline{\mathcal A}_{\Pi}(I)$, $\overline{\mathcal A}_{\Pi}(O)$, and $\overline{\mathcal A}_{\Pi}(P)$. 

\SPL\ was evaluated on
a dozen of \textsc{C} numerical programs collected from several different folders (categories) of the
8th International Competition on Software Verification (SV-COMP 2019, \url{https://sv-comp.sosy-lab.org/2019/})
as well as from the real-world BusyBox project (\url{https://busybox.net}).
The folders from SV-COMP we use are: \texttt{loops}, \texttt{loop-invgen} (\texttt{invgen} for short), \texttt{loop-lit} (\texttt{lit} for short), \texttt{termination-crafted} (\texttt{crafted} for short). 
In case of SV-COMP, we have first selected some numerical programs with integers, and then
we have manually added variability (features and \texttt{\#if} directives) in each of them.
In case of BusyBox, we have first selected some programs with numerical features, and then
we have simplified those programs so that our tool can handle them. For example, any reference to a pointer or a library
function is replaced with $[-\infty,+\infty]$.
Table~\ref{fig:performance1} presents characteristics of the selected
benchmarks. 
We list: the file name (Benchmark), folder where it is located (folder),
number of features ($|\Ff|$), number of configurations ($|\Kk|$), and number of
lines of code (LOC).

\paragraph*{Performance Results}
Table~\ref{fig:performance1}  shows the results of analyzing our benchmark files by using
different versions of our lifted static analyses based on decision trees and on tuples.
For each version of decision tree-based lifted analysis,
there are two columns.
In the first column, \textsc{Time}, we report the running time in seconds to analyze the given benchmark using the
corresponding version of lifted analysis based on decision trees.
In the second column, \textsc{Impr.}, we report the speed up factor for each version of lifted analysis based on decision trees relative to
the corresponding baseline lifted analysis based on tuples
($\overline{\mathcal A}_{\Tt}(I)$ vs. $\overline{\mathcal A}_{\Pi}(I)$, $\overline{\mathcal A}_{\Tt}(O)$ vs. $\overline{\mathcal A}_{\Pi}(O)$, and $\overline{\mathcal A}_{\Tt}(P)$ vs. $\overline{\mathcal A}_{\Pi}(P)$).
The performance results confirm that sharing is indeed effective and especially so for large values of $|\Kk|$.
On our benchmarks, it translates to speed ups (i.e., ($\overline{\mathcal A}_{\Tt}(-)$ vs. $\overline{\mathcal A}_{\Pi}(-)$) that
range from 1.1 to 4.6 times when $|\Kk|\!<\!100$, and from 3.7 to 32 times when $|\Kk|\!>\!100$.
Notice that $\overline{\mathcal A}_{\Tt}(I)$ is the fastest version, 
and $\overline{\mathcal A}_{\Tt}(P)$ is the slowest but the most precise.

\newcommand{\ra}[1]{\renewcommand{\arraystretch}{#1}}

\begin{table*}[t]\centering
\ra{1.03}
\caption{Performance results for lifted static analyses based on decision trees vs. tuples (which are used as baseline). 
All times are in seconds.
}\label{fig:performance1}
\begin{tabular}{@{}lllllllllll@{}}\toprule
\multirow{2}{*}{Benchmark} & \multirow{2}{*}{\small{folder}}  & \multirow{2}{*}{~$|\Ff|$~} & \multirow{2}{*}{~$|\Kk|$~} & \multirow{2}{*}{~\scriptsize{LOC}~} & \multicolumn{2}{c|}{$\overline{\mathcal A}_{\Tt}(I)$}  &  \multicolumn{2}{c|}{$\overline{\mathcal A}_{\Tt}(O)$}   & \multicolumn{2}{c}{$\overline{\mathcal A}_{\Tt}(P)$} \\
\cmidrule{6-7} \cmidrule{8-9} \cmidrule{10-11}
     &  & & & & \footnotesize{\textsc{Time}}  &  ~\footnotesize{~\textsc{Impr.}}~  &  \footnotesize{\textsc{Time}}  &  ~\footnotesize{\textsc{Impr.}}~  &  \footnotesize{\textsc{Time}}  &  ~\footnotesize{\textsc{Impr.}}~ \\ \midrule
\footnotesize{\texttt{half\_2.c}} & \scriptsize{\texttt{invgen}}& \footnotesize{2} & \footnotesize{36} & \footnotesize{60}  & \footnotesize{0.010}   & ~\footnotesize{2.4$\times$}~ & \footnotesize{0.017} & ~\footnotesize{3.5$\times$}~ & \footnotesize{0.022} & ~\footnotesize{4.6$\times$}~  \\
 \footnotesize{\texttt{heapsort.c}} & \scriptsize{\texttt{invgen}} & \footnotesize{2} & \footnotesize{36} & \footnotesize{60}  & \footnotesize{0.036}   & ~\footnotesize{2.2$\times$}~ & \footnotesize{0.226} & ~\footnotesize{1.1$\times$}~ & \footnotesize{0.191} & ~\footnotesize{2.0$\times$}~  \\
  \footnotesize{\texttt{seq.c}} & \scriptsize{\texttt{invgen}} & \footnotesize{3} & \footnotesize{125} &  \footnotesize{40}  & \footnotesize{0.039}   & ~\footnotesize{9.3$\times$}~ & \footnotesize{0.460} & ~\footnotesize{4.3$\times$}~ & \footnotesize{0.164} & ~\footnotesize{11$\times$}~  \\ 
 \footnotesize{\texttt{eq1.c}} & \scriptsize{\texttt{loops}} & \footnotesize{2} & \footnotesize{36} &  \footnotesize{20} & \footnotesize{0.015}  &  ~\footnotesize{3.4$\times$}~ & \footnotesize{0.049}  & ~\footnotesize{3.1$\times$}~ & \footnotesize{0.052}  &    ~\footnotesize{4$\times$}~      \\ 
 \footnotesize{\texttt{eq2.c}} & \scriptsize{\texttt{loops}} & \footnotesize{2} & \footnotesize{25} & \footnotesize{20} & \footnotesize{0.013} & ~\footnotesize{1.9$\times$}~  & \footnotesize{0.047}  & ~\footnotesize{1.3$\times$}~  & \footnotesize{0.040}  &   ~\footnotesize{1.9$\times$}~           \\ 
 \footnotesize{\texttt{sum01*.c}} & \scriptsize{\texttt{loops}} & \footnotesize{2} & \footnotesize{25} & \footnotesize{20} & \footnotesize{0.016} & ~\footnotesize{1.7$\times$}~  & \footnotesize{0.086}  & ~\footnotesize{1.5$\times$}~  & \footnotesize{0.062}  &   ~\footnotesize{2.2$\times$}~         \\ 
\footnotesize{\texttt{hhk2008.c}} & \scriptsize{\texttt{lit}} & \footnotesize{3} & \footnotesize{216} & \footnotesize{30} & \footnotesize{0.023}   & ~\footnotesize{10$\times$}~ & \footnotesize{0.153}  & ~\footnotesize{4.5$\times$}~ & \footnotesize{0.074}  &    ~\footnotesize{12.5$\times$}~        \\ 
\footnotesize{\texttt{gsv2008.c}} & \scriptsize{\texttt{lit}} & \footnotesize{2} &  \footnotesize{25} & \footnotesize{25}  & \footnotesize{0.013}   & ~\footnotesize{1.5$\times$}~ & \footnotesize{0.035} & ~\footnotesize{1.2$\times$}~ & \footnotesize{0.037} &   ~\footnotesize{2$\times$}~    \\ 
 \footnotesize{\texttt{gcnr2008.c}} & \scriptsize{\texttt{lit}} & \footnotesize{2} &  \footnotesize{25} &  \footnotesize{30}  & \footnotesize{0.021}   & ~\footnotesize{2$\times$}~ & \footnotesize{0.070} & ~\footnotesize{2.1$\times$}~ & \footnotesize{0.102} & ~\footnotesize{2.6$\times$}~  \\ 
 \footnotesize{\texttt{Toulouse*.c}} & \scriptsize{\texttt{crafted}} & \footnotesize{3} & \footnotesize{125} & \footnotesize{75} & \footnotesize{0.043}   & ~\footnotesize{6.1$\times$}~ & \footnotesize{0.259}  & ~\footnotesize{2.4$\times$}~ & \footnotesize{0.175}  &  ~\footnotesize{7.6$\times$}~    \\ 
 \footnotesize{\texttt{Mysore.c}} & \scriptsize{\texttt{crafted}} & \footnotesize{3} & \footnotesize{125} & \footnotesize{35} & \footnotesize{0.019}   & ~\footnotesize{3.7$\times$}~ & \footnotesize{0.090}  & ~\footnotesize{1.1$\times$}~ & \footnotesize{0.056}  &  ~\footnotesize{5.4$\times$}~    \\ 
 \footnotesize{\texttt{copyfd.c}} & \scriptsize{\texttt{BusyBox}} & \footnotesize{1} & \footnotesize{16} & \footnotesize{84} & \footnotesize{0.013}   & ~\footnotesize{3.9$\times$}~ & \footnotesize{0.041}  & ~\footnotesize{6.2$\times$}~ & \footnotesize{0.054}  &  ~\footnotesize{5.2$\times$}~    \\
  \footnotesize{\texttt{real\_path.c}} & \scriptsize{\texttt{BusyBox}} & \footnotesize{2} & \footnotesize{128} & \footnotesize{45} & \footnotesize{0.023}   & ~\footnotesize{14$\times$}~ & \footnotesize{0.077}  & ~\footnotesize{28$\times$}~ & \footnotesize{0.085}  &  ~\footnotesize{32$\times$}~    \\
\bottomrule
\end{tabular}
\end{table*}

\paragraph*{Computational tractability}

The tuple-based lifted analysis $\overline{\mathcal A}_{\Pi}(-)$ may become very slow
or even infeasible for very large configuration spaces $|\Kk|$.
We have tested the limits of $\overline{\mathcal A}_{\Pi}(P)$ and $\overline{\mathcal A}_{\Tt}(-)$.
We took a method, \texttt{test}$_n^k$(), which contains $n$
numerical features \texttt{A}$_1, \ldots, \texttt{A}_n$, such that each numerical feature $\texttt{A}_i$ has domain
 $\mathrm{dom}(\texttt{A}_i)=[0,k-1]=\{0, \ldots, k-1\}$.
The body of \texttt{test}$_n^k$() consists of $n$ sequentially composed  $\texttt{\#if}$-s
of the form $\texttt{\#if} \, (\texttt{A}_i=0) \ \impassign{\texttt{i}}{\texttt{i+}1} \ \texttt{\#else} \  \impassign{\texttt{i}}{0} \ \texttt{\#endif}$
For example, \texttt{test}$_2^3$() with two features \texttt{A}$_1$ and \texttt{A}$_2$,
whose domain is $[0,2]$, is:
\begin{center}
	$\begin{array}{ll}
\textcolor{gray}{{\ST 1}} \qquad & \texttt{int} \ \impassign{\texttt{i}}{0}; \\
\textcolor{gray}{{\ST 2}} \qquad &  \texttt{\#if} \, (\texttt{A}_1=0) \ \impassign{\texttt{i}}{\texttt{i+}1} \ \texttt{\#else} \  \impassign{\texttt{i}}{0} \ \texttt{\#endif} \\
\textcolor{gray}{{\ST 3}} \qquad &  \texttt{\#if} \, (\texttt{A}_2=0) \ \impassign{\texttt{i}}{\texttt{i+}1} \ \texttt{\#else} \  \impassign{\texttt{i}}{0} \ \texttt{\#endif} \ \textcolor{gray}{{\ST 4}}
	\end{array}$
\end{center}
Subject to the chosen configuration, the variable \texttt{i} in location \ST 4 can have
a value in the range from value 2 when \texttt{A}$_1$ and \texttt{A}$_2$ are assigned to 0, to value 0 when $\texttt{A}_2 \geq 1$.
The analysis results in location \ST 4 of \texttt{test}$_2^3$() obtained using $\overline{\mathcal A}_{\Pi}(P)$
and $\overline{\mathcal A}_{\Tt}(P)$ are shown in Fig.~\ref{fig:foo3:tuple} and Fig.~\ref{fig:foo3:dt}, respectively.
$\overline{\mathcal A}_{\Pi}(P)$ uses tuples with 9 interval properties (components), while
$\overline{\mathcal A}_{\Tt}(P)$ uses 3 interval properties (leafs). 

\begin{figure*}[t]
\begin{minipage}[b]{0.48\textwidth}
\centering
$\begin{array}{@{} l @{}}
{\scriptscriptstyle \big(\! \overbrace{[\texttt{i} = 2]}^{\texttt{A}_1=0 \land \texttt{A}_2=0}, \overbrace{[\texttt{i} = 0]}^{\texttt{A}_1=0 \land \texttt{A}_2=1},  \overbrace{[\texttt{i} = 0]}^{\texttt{A}_1=0 \land \texttt{A}_2=2}, } \\
{\scriptscriptstyle \overbrace{[\texttt{i} = 1]}^{\texttt{A}_1=1 \land \texttt{A}_2=0},
 \overbrace{[\texttt{i} = 0]}^{\texttt{A}_1=1 \land \texttt{A}_2=1}, \overbrace{[\texttt{i} = 0]}^{\texttt{A}_1=1 \land \texttt{A}_2=2}, } \\ {\scriptscriptstyle \overbrace{[\texttt{i} = 1]}^{\texttt{A}_1=2 \land \texttt{A}_2=0}, \overbrace{[\texttt{i} = 0]}^{\texttt{A}_1=2 \land \texttt{A}_2=1},
 \overbrace{[\texttt{i} = 0]}^{\texttt{A}_1=2 \land \texttt{A}_2=2}
 \!\! \big) }
	\end{array}$
\vspace{-1mm}
\caption{$\overline{\mathcal A}_{\Pi}(P)$ results at \ST 4 of \texttt{test}$_2^3$().}
\label{fig:foo3:tuple}
\end{minipage} \
\begin{minipage}[b]{0.48\textwidth}
\centering
\begin{tikzpicture}[-,>=stealth',level/.style={sibling distance = 1.25cm/#1,
  level distance = 0.5cm}]
     \node[arn_n](A){$\scriptstyle{\texttt{A}_2=0}$};
     \node[arn_n,below left=0.8cm and 0.2cm of A](B){$\scriptstyle{\texttt{A}_1=0}$};
     \node[arn_x,below right=0.8cm and 0.2cm of A](C){${\scriptstyle [\texttt{i} = 0]}$};
     \node[arn_x,below left=0.8cm and 0.2cm of B](G){${\scriptstyle [\texttt{i} = 2]}$};
     \node[arn_x,below right=0.8cm and 0.2cm of B](H){${\scriptstyle [\texttt{i} = 1]}$};
     \draw[dashed](A) to node[above right]{} (C);
     \draw[-](A) to node[above left]{} (B);
     \draw[-](B) to node[above left]{} (G);
     \draw[dashed](B) to node[above right]{} (H);
  \end{tikzpicture}
\vspace{-1mm}
\caption{$\overline{\mathcal A}_{\Dd}(P)$ results at \ST 4 of \texttt{test}$_2^3$().}
\label{fig:foo3:dt}
\end{minipage}
\end{figure*}

\begin{table*}\centering
\ra{1.08}
\caption{The performance results of analyzing \texttt{test}$_n^k$.
}\label{fig:foo:results}
\begin{tabular}{@{}llllllllll@{}}\toprule
~\multirow{2}{*}{\footnotesize n}~  & \multicolumn{3}{c|}{$k=3$} & \multicolumn{3}{c|}{$k=5$} & \multicolumn{3}{c}{$k=7$}  \\
\cmidrule{2-4} \cmidrule{5-7} \cmidrule{8-10}
   & {\footnotesize $\overline{\mathcal A}_{\Pi}(P)$} & {\footnotesize $\overline{\mathcal A}_{\Tt}(P)$} & ~{\footnotesize \textsc{Impr.}}~
& {\footnotesize $\overline{\mathcal A}_{\Pi}(P)$} & {\footnotesize $\overline{\mathcal A}_{\Tt}(P)$} & ~{\footnotesize \textsc{Impr.}}~
& {\footnotesize $\overline{\mathcal A}_{\Pi}(P)$} & {\footnotesize $\overline{\mathcal A}_{\Tt}(P)$} & ~{\footnotesize \textsc{Impr.}}~
 \\ \midrule
 ~5~   & 0.164   &  0.137  &  ~1.2$\times$~  & 2.859   &  0.139  &  ~20.6$\times$~  & 19.976   &  0.138  &  ~144.7$\times$~     \\
 ~6~   & 0.701   & 0.293   &  ~2.4$\times$~  & 23.224  &  0.294  &  ~79.1$\times$~  & {\tiny \texttt{infeasible}}   &  0.299  &  ~$\infty\times$~ \\
 ~8~   & 17.420   &  1.761  &  ~9.9$\times$~  & {\tiny \texttt{infeasible}}  &  1.765  &  ~$\infty\times$~ & {\tiny \texttt{infeasible}}   &  1.767  &  ~$\infty\times$~  \\
 ~10~  & 278.7   &  5.591  &  ~49.8$\times$~  & {\tiny \texttt{infeasible}}  &  5.596  &  ~$\infty\times$~  & {\tiny \texttt{infeasible}}   &  5.639  &  ~$\infty\times$~   \\
 ~11~  & {\tiny \texttt{infeasible}}   &  13.807  & ~$\infty\times$~   & {\tiny \texttt{infeasible}}  &  13.859  &  ~$\infty\times$~  & {\tiny \texttt{infeasible}}   &  13.809  &  ~$\infty\times$~ \\
 ~14~  & {\tiny \texttt{infeasible}}   &  327.10  &  ~$\infty\times$~  & {\tiny \texttt{infeasible}}  &  442.23  &  ~$\infty\times$~  & {\tiny \texttt{infeasible}}   &  459.19  &  ~$\infty\times$~ \\
\bottomrule
\end{tabular}
\end{table*}

We have generated methods \texttt{test}$_n^k$() by gradually increasing variability.
In general, the size of tuples used by $\overline{\mathcal A}_{\Pi}(P)$ is $k^n$,
whereas the number of leaf nodes in decision trees used by $\overline{\mathcal A}_{\Tt}(P)$
in the final program location is $n + 1$. 
The performance results of analyzing \texttt{test}$_n^k$, for different values of $n$ and $k$, using
$\overline{\mathcal A}_{\Pi}(P)$ and $\overline{\mathcal A}_{\Tt}(P)$ are shown in Table~\ref{fig:foo:results}.
In the columns \textsc{Impr.}, we report the speed-up of $\overline{\mathcal A}_{\Tt}(P)$
with respect to $\overline{\mathcal A}_{\Pi}(P)$.
We observe that $\overline{\mathcal A}_{\Tt}(P)$ yields decision trees that provide quite compact and symbolic representation of lifted analysis results.
Since the configurations with equivalent analysis results are nicely encoded using linear constraints in decision nodes,
the performance of $\overline{\mathcal A}_{\Tt}(P)$ does not depend on $k$, but only depends on $n$. On the other hand, the performance of $\overline{\mathcal A}_{\Pi}(P)$ heavily depends on $k$.
Thus, within a timeout limit of 300 seconds, the analysis $\overline{\mathcal A}_{\Pi}(P)$ fails to terminate for \texttt{test}$_{11}^3$, \texttt{test}$_8^5$, and \texttt{test}$_6^7$. 
In summary, we can conclude that decision trees $\overline{\mathcal A}_{\Tt}(P)$ can not only greatly speed up lifted
analyses, but also turn previously infeasible analyses into feasible. 

\vspace{-2mm}
\section{Related Work}

\emph{Decision-tree abstract domains} have been used in abstract interpretation community recently \cite{DBLP:conf/sas/GurfinkelC10,DBLP:conf/birthday/CousotCM10,DBLP:conf/sas/ChenC15,DBLP:conf/sas/UrbanM14}.
Decision trees have been applied for the disjunctive refinement of Interval domain \cite{DBLP:conf/sas/GurfinkelC10}.
That is, each element of the new domain is a propositional formula over interval linear constraints.
Segmented decision tree abstract domains has also been defined \cite{DBLP:conf/birthday/CousotCM10,DBLP:conf/sas/ChenC15}
to enable path dependent static analysis.
Their elements contain decision nodes that are determined either by values of program variables \cite{DBLP:conf/birthday/CousotCM10} or by
the branch (\texttt{if}) conditions \cite{DBLP:conf/sas/ChenC15}, whereas the leaf nodes are numerical properties.
Urban and Mine \cite{DBLP:conf/sas/UrbanM14} use decision tree-based abstract domains to prove program termination.
Decision nodes are labelled with linear constraints that split
the memory space and leaf nodes contain affine ranking functions for proving program termination.


Recently, two main styles of static analysis have been a topic of
considerable research in the SPL community: \emph{a dataflow
analysis from the monotone framework} developed by Kildall \cite{DBLP:conf/popl/Kildall73} that is algorithmically defined on syntactic CFGs, and \emph{an abstract interpretation-based static analysis} developed by Cousot and Cousot \cite{DBLP:conf/popl/CousotC77} that is more general and semantically defined.
Brabrand et. al.~\cite{taosd2013} lift a dataflow
analysis from the \emph{monotone framework}, resulting in a tuple-based lifted dataflow
analysis that works on the level of families. 
Another efficient
implementation of the lifted dataflow analysis from the monotone framework is based on using variational
data structures \cite{DBLP:journals/tosem/RheinLJKA18} (e.g., variational CFGs, variational data-flow facts).
Midtgaard et. al.~\cite{DBLP:conf/aosd/MidtgaardBW14} have proposed a formal methodology for systematic derivation of tuple-based lifted static
analyses 
in the \emph{abstract interpretation framework}.
A more efficient lifted static analysis by abstract interpretation obtained by improving representation via BDD domains is given in \cite{DBLP:conf/gpce/Dimovski19}.
Another approach to speed up lifted analyses is by using so-called variability abstractions \cite{ecoop15}, which
are used to derive abstract lifted analyses.
They tame the combinatorial explosion of the number of configurations and reduce it to something more tractable by
manipulating the configuration space. 
However, the above lifted analyses
are applied to program families with only Boolean features.
On the other hand, here we consider C families with both Boolean
and numerical features, which represent the majority of industrial embedded code.

\vspace{-2mm}
\section{Conclusion}

In this work we employ decision trees and widely-known numerical abstract domains for automatic inference
of invariants in all locations of C program families that contain numerical features.
In future, we would like to extend the lifted abstract domain to also support
non-linear constraints \cite{granger} and more complex heap-manipulating families \cite{DBLP:journals/corr/ChangR13}.
An interesting direction for future work would be to explore possibilities of applying variability abstractions \cite{ecoop15} 
as yet another way to speed up lifted analyses.



 \bibliographystyle{plain}
 \bibliography{mainBib}


\appendix
\newpage
\section{Appendix} \label{app:alg}

\begin{proof}[of Theorem~\ref{thm:1}]
The proof is by induction on the structure of $s$.
Assume $\gamma_{\Tt}(t) = \overline{\gamma}(\overline{a})$ (*).
We consider the two most interesting cases.
\begin{description}
\item[Case $\texttt{x:=}e$.]
$\overline{\textrm{ASSIGN}}(\overline{a}, \, \texttt{x:=}e)$
applies $\texttt{ASSIGN}_{\Aa}(t,\texttt{x:=}e)$ to each component of $\overline{a}$.
 On the other hand, \texttt{ASSIGN}$_{\Tt}(t,\texttt{x:=}e$) applies $\texttt{ASSIGN}_{\Aa}(t,\texttt{x:=}e)$
to each leaf $a$ in $t$.
The proof follows by correctness of the assumption (*).

\item[Case $\texttt{\#if}\, (\theta)\, s~\texttt{\#endif}$.]
Transfer functions for \texttt{\#if} are identical in both lifted domains.
We only need to show that $\overline{\textrm{FEAT-FILTER}}(\overline{a}, \, \theta)$
and \texttt{FEAT-FILTER}$_{\Tt}(t,\theta$) are identical.
This can be shown by induction on $\theta$.
Assume that $\theta$ is an atomic constraint.
$\overline{\textrm{FEAT-FILTER}}(\overline{a}, \, \theta)$  keeps only those
components $k$ of $\overline{a}$ such that $k \models \theta$.
On the other hand, \texttt{FEAT-FILTER}$_{\Tt}(t,\theta$) first produces
all linear constraints in $\Tt$ that satisfy $\theta$, and then adds them
in the tree $t$. Thus, it keeps only those leaf nodes that satisfy
the newly generated constraints from $\theta$. The other cases are similar.
\end{description}
\end{proof} 


\end{document}